\documentclass[twocolumn,superscriptaddress,showpacs,prd,
aps,amsmath,amssymb,nofootinbib,eqsecnum]{revtex4-1}
\usepackage{hyperref}
\usepackage{graphicx}
\usepackage{bm} 
\usepackage{mathrsfs} 
\usepackage[normalem]{ulem} 
\usepackage{multirow}

\usepackage{color}
\usepackage{comment}            

\definecolor{mblue}{rgb}{0,0.2,1}
\definecolor{mgreen}{rgb}{0.1,0.8,0.1}
\definecolor{violet}{rgb}{0.5,0,0.5}
\definecolor{orange}{rgb}{0.9,0.3,0}

\newcommand{\be}{\begin{equation}}
\newcommand{\ee}{\end{equation}}
\newcommand{\bea}{\begin{eqnarray}}
\newcommand{\eea}{\end{eqnarray}}
\newcommand{\M}{\mathscr{M}}

\newcommand{\vv}[1]{\bm{#1}}

\newcommand{\Lie}[1]{\mathcal{L}_{#1}\,}

\newcommand{\eps}{\epsilon}

\newcommand{\cD}{\mathcal{D}}
\newcommand{\cH}{\mathcal{H}}

\newcommand{\bam}{\textsc{bam }}
\newcommand{\sgrid}{\textsc{sgrid }}

\usepackage{amsmath}
\usepackage{amssymb}

\begin{document}

\title{Initial data for binary neutron stars with adjustable eccentricity}

\author{Niclas Moldenhauer}
\email{niclas.moldenhauer@uni-jena.de}
\affiliation{
Theoretical Physics Institute, University of Jena, Max-Wien-Platz 1, 07743 Jena, Germany}

\author{Charalampos M.\ Markakis} 
\email{charalampos.markakis@uni-jena.de}
\affiliation{
Theoretical Physics Institute, University of Jena, Max-Wien-Platz 1, 07743 Jena, Germany}
\affiliation{
Mathematical Sciences, University of Southampton, Southampton, SO17 1BJ, United Kingdom}

\author{Nathan K.\ Johnson-McDaniel}
\email{nathan-kieran.johnson-mcdaniel@uni-jena.de}
\affiliation{
Theoretical Physics Institute, University of Jena, Max-Wien-Platz 1, 07743 Jena, Germany}

\author{Wolfgang Tichy}
\email{wolf@fau.edu}
\affiliation{Physics Department, Florida Atlantic University, 777 Glades Road, Boca Raton, FL 33431, USA
}

\author{Bernd Br\"ugmann}
\email{bernd.bruegmann@uni-jena.de}
\affiliation{
Theoretical Physics Institute, University of Jena, Max-Wien-Platz 1, 07743 Jena, Germany}

\date{\today}  

\begin{abstract} 
Binary neutron stars in circular orbits can be modeled as helically symmetric, i.e., stationary in a rotating frame.
This symmetry gives rise to  a first integral of the Euler equation, often employed  for constructing equilibrium solutions via iteration.
For eccentric orbits, however, the lack of helical symmetry has prevented the use of this method, and the numerical relativity community has often resorted to constructing initial data by superimposing boosted spherical stars without solving the Euler equation.
The spuriously excited neutron star oscillations seen in evolutions of such data
arise because such configurations lack the appropriate tidal deformations
and are  stationary in a linearly comoving---rather than rotating---frame.
We consider eccentric configurations at apoapsis that are instantaneously stationary in a rotating frame. We extend the notion of  helical symmetry to eccentric orbits, by approximating the elliptical orbit of each companion as instantaneously circular,  using the ellipse's inscribed circle.  The      two inscribed helical   symmetry vectors  give rise to approximate instantaneous  first integrals of the Euler equation throughout each companion. We use these integrals as the basis of a self-consistent iteration of the Einstein constraints to construct conformal thin-sandwich initial data for eccentric binaries. We find that the spurious stellar oscillations are  reduced by at least an order of magnitude, compared with those found in evolutions of superposed initial data. The tidally induced oscillations, however, are physical and qualitatively similar to earlier evolutions. Finally, we show how to incorporate radial velocity due to radiation reaction in our inscribed helical symmetry vectors, which would allow one to obtain truly non-eccentric initial data when our eccentricity parameter $e$ is set to zero.

\end{abstract} 


\maketitle

\section{Introduction}
\label{s:intro}

The coalescences of binary neutron stars are a prominent source for ground-based gravitational wave detectors. (See Table~1 in~\cite{FabRas12} for a review of population synthesis predictions and~\cite{DomBerOSh14} for more recent work.) In particular, binary neutron stars are the only compact binary sources relevant for ground-based detectors that have been observed to date (via electromagnetic observations of binary pulsars). Neutron star binary coalescences are also interesting beyond gravitational wave astronomy as potential progenitors of short gamma-ray bursts and sources of the $r$-process material that enriches the interstellar medium with heavy elements~\cite{EicLivPir89}. 

All the known binary neutron stars are at least somewhat eccentric (eccentricities between $0.085$ and $0.681$ for the systems that will merge within a few hundred million years; see Tables~2 and~3 in~\cite{PosYun14}), though these will be highly circular when they merge, since gravitational radiation reaction efficiently circularizes the orbit~\cite{Pet64}; see Table~1 in~\cite{KowBulBel11}. However, there are possible (though likely rare) scenarios in which neutron stars can merge with nonnegligible eccentricity, either because they are formed 
with a high eccentricity and small periapsis distance by 
dynamical interactions in dense stellar regions, such as globular clusters~\cite{OLeKocLoe09,LeeRuiVen09,Tsa13,SamMacRam13} or have their eccentricity excited by, e.g., the Kozai mechanism in a hierarchical triple~\cite{AntPer12,NaoKocLoe12,Set13,AntShaTho14} (\cite{AntMurMik13} also treats the latter case, but only considers the case of binary black holes). Gravitational waves from highly eccentric compact binary systems exhibit a repeated burst structure, which poses challenges for gravitational wave astronomy, but also offers potential rewards, as has been explored in a number of works:
McWilliams, Pretorius, and collaborators~\cite{EasMcWLev13, TaiMcWPre14} discuss strategies for detection, while Kyutoku and Seto~\cite{KyuSet13} find improvements in the accuracy of premerger sky localization and timing, compared to the quasicircular case. Loutrel, Yunes, and Pretorius~\cite{LouYunPre14} consider bursts from highly eccentric binaries as a regime for testing general relativity, while Tsang
~\cite{Tsa13} has considered the possibility of obtaining electromagnetic flares from crust cracking during close encounters.

The first full numerical relativity evolutions of highly eccentric binary neutron stars were carried out by Gold~\emph{et al.}~\cite{GolBerThi11}, with a further study by East and Pretorius~\cite{EasPre12}. These systems have also been simulated with Newtonian methods in~\cite{LeeRuiVen09,RosPirNak12}. 
In addition, there have been full numerical relativity simulations of highly eccentric black hole-neutron star (BHNS) binaries~\cite{SteEasPre11,EasPreSte12} and binary black holes, e.g.,~\cite{GolBru09,GolBru12}.
However, all the evolutions of highly eccentric binaries with neutron stars have used inconsistent initial data, due to the difficulty of generalizing the standard procedure for quasicircular orbits, where one uses the binary's approximate helical Killing vector to solve the Euler equation via a first integral. In particular, Gold~\emph{et al.}~\cite{GolBerThi11} used a superposition of boosted spherical stars, which leads to relatively large constraint violations in addition to not giving the appropriate matter configuration. The Princeton group~\cite{EasPre12,SteEasPre11,EasPreSte12} solves the constraints to obtain their initial data, as described in~\cite{EasRamPre12}, but they do not solve the Euler equation. It would obviously be desirable to obtain consistent initial data for these configurations.
In particular, it is possible that the tidally induced oscillations of the neutron star(s) found in these evolutions are affected by the initial spurious oscillations of the neutron star from imperfect initial data.

Here we present a method to construct binary neutron star initial data with arbitrary eccentricity, by generalizing the helical Killing vector to a pair of inscribed   helical symmetry vectors, appropriate for the more general situation of an eccentric orbit at apoapsis. We then provide
a first proof-of-principle numerical implementation of this method for equal-mass binaries; our method is applicable  to arbitrary mass ratios or BHNS binaries as well. This self-consistent method yields constraint-solved initial data (in the Isenberg-Wilson-Mathews approximation~\cite{Ise08,Wilson1989}, i.e., assuming spatial conformal flatness), where both the geometry and matter are momentarily stationary in a rotating frame. We also give an additional generalization to include radial velocity, though we do not implement this numerically in this paper. As a test of the method, we show that it produces results with the expected physical properties both in the quasicircular limit, where we compare with the results of data calculated with the standard method, as well as for nonzero values of the eccentricity.

For simplicity, we have considered irrotational binaries in the current numerical implementation and made the approximation of a homogeneous velocity field. However, it is possible (and relatively straightforward) to drop these approximations and even add spin to the construction, following~\cite{Tic11,Tic12}. In particular, the assumption of a homogeneous velocity field is made merely for convenience, so that we can use a Cartesian multigrid elliptic solver without surface-fitted coordinates, but we demonstrate that it is a reasonable approximation for sufficiently separated binaries.
The assumption of irrotational stars is standard and is reasonable for a first study: While neutron stars can spin quite quickly (at least up to $716$~Hz~\cite{HesRanSta06}), the spins in known binary neutron stars are much more modest, at most $44$~Hz for the more massive 
star in the double pulsar (see, e.g., Table~2 in~\cite{PosYun14}), and all of these will decrease further due to spin-down before the stars merge.
Moreover, the viscosity of neutron star matter is far too low for the stars to experience any significant tidal spin-up, as established by Kochanek~\cite{Koc92} and Bildsten and Cutler~\cite{BilCut92}. Thus, it has been standard to consider irrotational flow in modeling binary neutron stars (see, e.g.,~\cite{FabRas12}), since the system's orbital frequency ten to twenty orbits before merger is $\gtrsim 100$~Hz in the quasicircular case.

However, the extent to which the relatively small spins of the members of observed binary neutron stars is a selection effect remains unclear, particularly because the known population is so small (only 9 systems~\cite{PosYun14}). Moreover, if one did form a binary with even a modestly rapidly spinning neutron star, the spin-down before merger might not be very significant: Fast-spinning neutron stars are thought to have had their spin increased by accretion (a process known as recycling), which also reduces their external magnetic field, and thus reduces the stars' spin-down, as well (see, e.g.,~\cite{Tau11}). In particular, \cite{Tic11} finds that the more massive star in the double pulsar is expected to have spun down only to $37$~Hz at 
merger, and \cite{BerDieTic13} has found that spins of about this magnitude can have a sizable effect on the system's dynamics.

The addition of spin may be particularly interesting for eccentric systems,
since one way of forming such binaries is through dynamical assembly in dense
stellar systems, such as globular clusters, and globular clusters are a
fertile breeding ground for millisecond pulsars, including the fastest pulsar known.
For instance, all $23$ known pulsars in the rich globular cluster 47~Tucanae have
spin frequencies greater than $125$~Hz, and all but three are above $200$~Hz~\cite{Fre13}.

The paper is structured as follows: We first review the necessary portions of perfect fluid hydrodynamics and the $3+1$ split of the Einstein equations in Secs.~\ref{s:fluid} and~\ref{s:grav}, before describing the specific construction we use to obtain an approximate first integral to the Euler equation in Sec.~\ref{s:statfluidapprox}. We then describe the numerical implementation of the method in Sec.~\ref{s:num_meth} and evaluate its performance in the quasicircular case in Sec.~\ref{s:quasicirc} before giving examples of eccentric binaries in Sec.~\ref{s:ecc}. We discuss and conclude in Sec.~\ref{s:concl}, and give some ancillary results for nonrelativistic incompressible binaries in the Appendix.

We use the following notation throughout: We use Greek letters $\alpha,\beta,\gamma,\delta,...$ and $\mu,\nu,\kappa,\lambda,...$ for abstract and concrete spacetime indices, respectively. We also use Latin letters $a,b,c,...$ and $i,j,k,...$ for abstract and concrete spatial indices, respectively. We raise and lower concrete spatial indices with the flat conformal metric, while all other indices are raised and lowered with the physical metric;
the summation convention is always in force. We shall also use index-free notation when convenient, denoting vectors (spatial or spacetime) using boldface. We employ units with $G = c = M_\odot =1$
almost exclusively, except that we show the appearances of $G$ explicitly for clarity when making some Newtonian calculations in Sec.~\ref{s:eccorbitshelsym}. 

\section{Perfect fluid model}
\label{s:fluid}

\subsection{Thermodynamic quantities}

We consider a spacetime $(\M,g_{\alpha \beta})$, i.e., a manifold $\M$ endowed with a Lorentzian metric $g_{\alpha\beta}$. Furthermore, we assume that this spacetime is globally hyperbolic,
so it possesses a Cauchy surface (and, indeed, can be foliated by Cauchy surfaces).
Part of this spacetime is occupied by
a perfect fluid,
characterized by the  
energy-momentum tensor
\be \label{e:energymomentumfluid}
T_{\alpha \beta} = (\eps + p) \, u_\alpha u_\beta + p g_{\alpha \beta} ,
\ee
where $u^\alpha$ is the four-velocity, $\eps$ is the proper energy density, and $p$ the fluid pressure. 
Moreover, we assume that the fluid is a \emph{simple fluid}, i.e., that all the thermodynamic quantities depend only on the entropy density $\sigma$ and on the 
proper baryon number density $n$. In particular, the relation
\be \label{e:EOS}
  \eps = \eps(\sigma,n) . 
\ee
 is called the equation of state (EOS)  of the fluid. 
The temperature $T$ and the baryon chemical potential $\mu$ are then defined by
\be \label{e:def_T_mu}
  T := \frac{\partial\eps}{\partial\sigma} \qquad\mbox{and}\qquad
  \mu := \frac{\partial \eps}{\partial n} .
\ee
Then, the first law of thermodynamics can be written as
\be \label{e:deps1stlaw}
d\eps =\mu \,dn+Td\sigma.
\ee
As a consequence, $p$ is a function of $(\sigma,n)$ 
entirely determined by (\ref{e:EOS}): 
\be \label{e:p_EOS}
  p = -\eps + T  \sigma + \mu n . 
\ee
Let us introduce the specific enthalpy, 
\be \label{e:def_h}
  h := \frac{\eps+p}{\rho} = \frac{\mu}{m_{\rm b} \, } + T s ,
\ee
where $m_{\rm b} = 1.66\times 10^{-27} {\rm\; kg}$ (the atomic mass unit, which can be taken as the average nucleon mass) is
the baryon rest mass, 
$\rho$ is the rest-mass density
\be 
  \rho := m_{\rm b} \, n,
\ee
and  $s$ is the specific entropy:
\be \label{e:def_S}
  s := \frac{\sigma}{\rho} .
\ee
The second equality in (\ref{e:def_h}) is an immediate consequence of (\ref{e:p_EOS}). From Eqs.~\eqref{e:deps1stlaw}--\eqref{e:def_S},
we obtain the  thermodynamic relations \cite{Gou06,AndCom07}
\be \label{eq:depsdpthermodynamic}
d\eps =hd\rho+\rho T ds, \quad dp=\rho(dh-Tds).
\ee

To describe the matter inside a neutron star, we have to make a choice for the EOS in order to close the system. For the present study, we restrict attention to polytropes. Specifically, we assume that
  \be
  p = \kappa \rho^{1+1/n},
  \ee
 where $\kappa$ is the polytropic constant and  $n = \frac{1}{\Gamma-1}$ is the polytropic index ($\Gamma$ is the adiabatic index). Then, one may express $\rho$, $p$, and $\eps$ as functions of the specific enthalpy $h$
 \begin{subequations}
\begin{align}
 \rho &= \left[\frac{h-1}{\kappa (1+n)}\right]^n, \\
 p &= \kappa \left[\frac{h-1}{\kappa (1+n)}\right]^{1+n}, \\
 \eps &= \left[1+\frac{n(h-1)}{1+n}\right] \left[\frac{h-1}{\kappa (1+n)}\right]^n.
\end{align}
\end{subequations}
For simplicity, we shall only consider the case $n = 1$ (i.e., $\Gamma = 2$) in the numerical examples of this paper, since this is close to the effective polytropic index of realistic nuclear physics equations of state~\cite{LatPra00}.
Treatment of realistic  EOSs is straightforward using a piecewise polytropic approximation \cite{ReaLacOwe09}.

\subsection{Euler equation}

The relativistic Euler equation follows from the conservation law of energy-momentum: 
\be \label{e:cons_enermom}
  \nabla_\alpha T^\alpha_{\,\,\,\,\beta}  = 0 , 
\ee
where $\nabla_\alpha$ denotes the  covariant derivative compatible with the metric $g_{\alpha \beta}$.
Using Eqs.~\eqref{e:deps1stlaw}--\eqref{eq:depsdpthermodynamic}, the divergence 
 of the  fluid energy-momentum tensor \eqref{e:energymomentumfluid} can be decomposed as
\be \label{eq:fluidenmomediv}
\nabla_\alpha T^\alpha_{\,\,\,\,\beta}   =\pi_\beta\nabla_\alpha (\rho u^\alpha)+ 
  \rho[ u^{\alpha} ({\rm{d}} \pi)_{\alpha \beta} - T \nabla_\beta  s ].
\ee
Here, $\pi_\alpha=h u_\alpha$ denotes the canonical momentum 1-form of a fluid element, while its exterior derivative 
$({\rm{d}} \pi)_{\alpha \beta}=\nabla_\alpha {\rm{}} \pi_\beta-\nabla_\beta {\rm{}} \pi_\alpha$ denotes  the canonical vorticity 2-form. Invoking the baryon number conservation law 
\be   \label{eq:continuityeqn}
\nabla_\alpha (\rho u^\alpha)=0,
\ee
Eqs.~\eqref{e:cons_enermom} and \eqref{eq:fluidenmomediv} yield the Euler equation,
\bea
    u^{\alpha} ({\rm{d}} \pi)_{\alpha \beta} = T \nabla_\beta  s,  \label{e:MHD-Euler}
\eea
written here in the Carter-Lichnerowicz form 
\cite{Syn37,Lic41,Tau59,Car79,Gou06}.
This particular form 
is quite useful when the fluid configuration possesses  certain symmetries \cite{GouMarUry11}. 
In the ``dust'' limit ($h\rightarrow1$, $T\rightarrow0$), Eq.~(\ref{e:MHD-Euler}) reduces to the equation satisfied by a pressure-free perfect fluid, i.e., the geodesic equation.
Projecting the Euler equation along $u^\beta$ yields
\bea
  && u^\beta \nabla_\beta s = 0,  \label{e:S_conserved} 
\eea
 implying that specific entropy is constant along flow lines. 

If the fluid is  \textit{barotropic},  
the thermodynamic quantities depend only on the 
proper baryon number density $n$ (or, equivalently,  rest mass density $\rho$). Particular examples are  cold ($T=0$) or  homentropic ($s=\rm{const.}$) fluids.
This assumption is  appropriate for  inspiralling   neutron-star binaries, as shock heating is absent and the fluid temperature is much lower than the Fermi  temperature \cite{FriedmanStergioulas2013}. 
In the remainder of this paper,  we shall restrict our attention to  barotropic flows, for which the Euler equation~(\ref{e:MHD-Euler}) simplifies to 
\bea  \label{eq:EulerEqBarotropic}
    u^{\alpha} ({\rm{d}} \pi)_{\alpha \beta} = 0.  \label{e:cold-Euler}
\eea
Barotropic fluid streamlines are  \textit{geodesics} of a   Riemannian manifold with  metric $h^2 g_{\alpha \beta}$ \cite{Lichnerowicz1967}. Indeed, the  fluid element action
  \be \label{eq:fluidaction}
S= \int_{{\tau _1}}^{{\tau _2}} {L(x,u) d\tau }=  - \int_{{\tau _1}}^{{\tau _2}} {h(x)\sqrt { - {g_{\alpha \beta }}(x){u^\alpha }{u^\beta }} d\tau } 
\ee
can be minimized (most easily via covariant techniques \cite{Car79}) to obtain  the  Euler equation  \eqref{e:cold-Euler}, with  canonical velocity given by $u^\alpha=dx^\alpha/d\tau$ and canonical momentum given by  $\pi_\alpha=\partial L/\partial u^\alpha = h u_\alpha$. 

\subsection{Noether's theorem \& Bernoulli's principle}

If   the 
 $\varepsilon$-family of infinitesimal coordinate
transformations $x^\alpha \rightarrow x^\alpha+ \varepsilon \, k^\alpha $ is a
continuous symmetry of 
 the
 fluid element action \eqref{eq:fluidaction},
then Noether's theorem implies that the quantity
\be \label{eq:injectionEnergyNoether}
\mathcal{E}=-k^\alpha \pi_\alpha
\ee
is conserved along streamlines. Indeed, using the equation of motion 
\eqref{eq:EulerEqBarotropic} and the constraint $u_\alpha u^\alpha=-1$, one finds
\be \label{eq:injectionEnergyBernoulli1}
u^\alpha \nabla _\alpha\mathcal{E} =- \frac{1}{{2h}}{{u^\alpha }{u^\beta }\Lie{\vv k}}({h^2}{g_{\alpha \beta }})=0,
\ee
where  $\Lie{\vv k}$ denotes the Lie derivative with respect to the vector $\vv k$. In geometrical terms, this result follows from the fact  that $k^\alpha$ is a Killing vector of the conformal metric $h^2 g_{\alpha \beta}$.
The conservation of the quantity~\eqref{eq:injectionEnergyNoether} is a generalization of Bernoulli's principle,  which is recovered in the Newtonian limit  if the Killing vector  generates time translations that leave the flow unchanged (i.e., if the flow is stationary) \cite{Gou06}. Note, however, that a Killing symmetry only guarantees a \textit{weak} Bernoulli principle, in the sense that the quantity  \eqref{eq:injectionEnergyNoether} is  conserved only along streamlines, but could differ from one  streamline to the next.
In order to obtain a \textit{strong} Bernoulli principle, i.e., a quantity conserved throughout the fluid,  a second condition (such as irrotationality or rigidity) is required. This issue will be revisited  in Sec.~\ref{s:statfluidapprox}.

\section{Gravitational field equations}
\label{s:grav}

\subsection{Extended conformal thin-sandwich formulation}

We consider a spacetime
$\M = \mathbb{R} \times \Sigma$ which is foliated
by a family of spacelike surfaces $\Sigma_t$.
 Making the standard $3+1$ decomposition in a chart $\{t,x^i\}$, the spacetime metric takes the form
\begin{equation}
ds^2=-\alpha^2 dt^2 +  \gamma_{ij}(dx^i+\beta^i dt)(dx^j+\beta^j dt),
\end{equation}
where $\alpha$ is the lapse  and $\beta^\alpha$ is the shift vector. These are related to the unit normal $n^\alpha$ of the three dimensional spatial hypersurface $\Sigma_t$ and the timelike vector $t^\alpha$ via $t^\alpha=\alpha n^\alpha+\beta ^\alpha$. The shift is purely spatial and satisfies $\beta^\alpha n_\alpha=0$, while we define the spatial metric $\gamma_{ab}(t)$ by restricting the projection tensor $\gamma_{\alpha\beta} = g_{\alpha\beta}+n_\alpha n_\beta$ to $\Sigma_t$.

In the extended conformal thin-sandwich (XCTS)\ formulation  \cite{BauCooSch98,Yor99,UryEri00,PfeYor02,HuaMarSug08,Ise08,BauSha10}  one decomposes the spatial metric into a spatial conformal metric $\bar \gamma_{ab}$  and a conformal factor $\psi$ defined by $\gamma_{ab}=\psi^4 \bar \gamma_{ab}$. One also decomposes the extrinsic curvature of the foliation as
\begin{equation}
K_{ab}:=-\frac{1}{2}\Lie{\vv n}\gamma_{ab}=A_{ab}+\frac{1}{3}\gamma_{ab}K,
\end{equation}
where
$K := K^a{}_a$ is the trace of the extrinsic curvature and
\begin{equation}
\label{e:Aab}
 A_{ab}=-\frac{\psi^4}{2\alpha}\left[\partial_t \bar\gamma_{a b} -(\bar L \beta)_{ab}\right]
\end{equation}
is the rescaled traceless part of the extrinsic curvature. Here,
$\partial_\mu$ denotes a partial derivative with respect to the coordinate $x^\mu$ on $\M$, while
\begin{equation}
(\bar L \beta)_{ab}=\bar D_a \beta_b+\bar D_b \beta_a-\frac{2}{3} \bar\gamma_{ab} \bar D_c \beta^c 
\end{equation}
is the traceless part of $\Lie{\vv \beta} \bar \gamma_{ab}$ and $\bar D_a$ is the covariant derivative  compatible with  $\bar \gamma_{ab}$.

Decomposing Einstein's equations $G_{\alpha\beta} = 8\pi T_{\alpha\beta}$ (where $G_{\alpha\beta}$ is the Einstein tensor), following the XCTS formulation, we take a set of five equations
\begin{subequations}
\begin{align}
          \label{eq:cts_ham1} 
          (G_{\alpha \beta}-8 \pi T_{\alpha \beta})n^\alpha n^\beta&=0,\\
          \label{eq:cts_mom1} 
          (G_{\alpha \beta}-8 \pi T_{\alpha \beta})\gamma_c{}^\alpha n^\beta&=0,\\
          \label{eq:cts_alphapsi1}
           (G_{\alpha \beta}-8 \pi T_{\alpha \beta})\left( \gamma^{\alpha \beta}+\frac{1}{2}n^\alpha n^\beta \right)&=0,
\end{align}
\end{subequations}
and solve them for the five metric coefficients $\{\psi,\alpha,\beta^a \}$ on the initial slice $\Sigma_0$.
The constraint equations \eqref{eq:cts_ham1} and \eqref{eq:cts_mom1}, along with \eqref{eq:cts_alphapsi1} can be written  in the form of  elliptic equations
with  nonlinear source terms:
\begin{subequations}
\label{eq:cts}
\begin{align}
 \label{eq:cts_ham} 
          \bar D^2\psi&=\frac{\psi}{8}\bar R-\frac{\psi^5}{8}\left(A_{ab}A^{ab}-\frac{2}{3}K^2+16\pi \rho_{\rm H} \right),  \\
          \bar\Delta_L\beta^a &= (\bar L \beta)^{ab}\bar D_b \ln(\alpha\psi^{-6}) + 16\pi\alpha\psi^{4}J^a\nonumber\\
          &\quad- \alpha\psi^{-6}\bar D_b(\alpha^{-1}\psi^{6} \partial_t \bar\gamma^{ab})+\frac{4}{3}\alpha \bar D^a K ,  \label{eq:cts_mom} \\ 
          \bar D^2(\alpha\psi)&=\alpha\psi^{5}\left[\frac{7}{8}A_{ab} A^{ab}+\frac{5}{12}K^2+2\pi(\rho_{\rm H}+2J)\right] \nonumber\\
        &\quad-\psi^5(\partial_t -\beta^b \bar D_b )K + \frac{1}{8}\alpha\psi \bar{R}, \qquad 
         \label{eq:cts_alphapsi}
\end{align}
\end{subequations}
where $\bar{D}^2 := \bar{D}^a\bar{D}_a$, $\bar{R}$ is the Ricci scalar of the spatial conformal metric ($\bar{R}_{ab}$ is its Ricci tensor) and $\bar{\Delta}_L\beta^a := \bar{D}^2\beta^a + \frac{1}{3}\bar{D}^a(\bar{D}_b\beta^b) + \bar{R}^a{}_b\beta^b$.

 The matter source terms are $\rho_{\rm H}$, $J^i$, and $J$, which correspond to the energy density, the momentum flux, and the
trace of the stress tensor. They are defined as projections of the
stress-energy tensor $T_{\alpha\beta}$ and thus can be written as
\begin{subequations}
\label{eq:sourceterms_orig}
\begin{align}
 \rho_{\rm H}&=T_{\alpha\beta}n^\alpha n^\beta, \\
 J^c &= -T_{\alpha\beta}\gamma^{c\alpha} n^\beta,\\
 J &= T_{\alpha\beta}\gamma^{\alpha\beta}.
\end{align}
\end{subequations}
Using the perfect fluid stress-energy tensor~\eqref{e:energymomentumfluid} and our assumption of spatial conformal flatness, we obtain
\begin{subequations}\label{eq:sourceterms}
\begin{align}
 \rho_{\rm H}&=\rho h (\alpha u^t-1)^2+\epsilon, \\
 J^i &= \rho h \,\alpha (u^t)^2 \,\psi^4 ( \beta^i + u^i/u^t), \\
 J &= \rho h [(\alpha u^t)^2-1]+3p.
\end{align} 
\end{subequations}
 If one imposes maximal slicing
\begin{subequations} 
 \label{eq:maximalslicing}
  \begin{align}
    \label{eq:maximalslicing1}
 K&=0,\\
   \label{eq:maximalslicing2}
 \partial_t K&=0,
\end{align}
\end{subequations}
  and assumes  spatial conformal flatness (Isenberg-Wilson-Mathews [IWM] approximation \cite{Ise08,Wilson1989})
\begin{subequations}
\label{eq:IWM}
 \begin{align}
  \label{eq:IWM1}
 \bar{\gamma}_{ab}&= f_{ab}, \\
\label{eq:IWM2} 
 \partial_t \bar{\gamma}_{ab}&=0,
\end{align}
\end{subequations}
(where $f_{ab}$ is the metric of flat space)
  and preserves these conditions in time (at least for an infinitesimally small time interval), then the XCTS equations \eqref{eq:cts} simplify considerably. In Cartesian coordinates $(f_{ij}=\delta_{ij})$, they reduce to \cite{BauSha10}       
\begin{subequations}
\label{eq:cts2}
\begin{align}
          \label{eq:cts_ham2} 
          \partial^i \partial_i\psi &=-\frac{1}{8}\psi^{5}(A_{ij}A^{ij}+16\pi\rho_{\rm H}),\\
          \label{eq:cts_mom2} 
          \partial^j \partial_j \beta^i
          +\frac{1}{3}\partial^i \partial_j \beta^j &=2\psi^{10} A^{ij}\partial_j (\alpha \psi^{-6})+16 \pi \alpha \psi^4 J^i,\\
          \partial^i \partial_i(\alpha\psi) &=\alpha\psi^5\left[\frac{7}{8}A_{ij} A^{ij}+2\pi(\rho_{\rm H}+2J)\right].
         \label{eq:cts_alphapsi2}
\end{align}
\end{subequations}
Here we raise and lower indices with the flat conformal metric, and will do so for all other concrete spatial indices.

 Note that, in the literature (as reviewed in, e.g., \cite{BauSha10}), authors traditionally invoke a time-like Killing symmetry or quasi-equilibrium to justify the conditions \eqref{eq:maximalslicing2}
and \eqref{eq:IWM2}. Typically, the maximal slicing and  spatial conformal flatness
 conditions   \eqref{eq:maximalslicing1} and \eqref{eq:IWM1} are imposed at a later stage. However, swapping the order of assumptions
makes the Killing symmetry redundant. That is,
if one imposes the conditions   \eqref{eq:maximalslicing1}, \eqref{eq:IWM1} from the beginning and preserves these conditions in time, then Eqs.~\eqref{eq:maximalslicing2},
 \eqref{eq:IWM2} follow \textit{without} assuming Killing symmetry or quasi-equilibrium. A notion of stationarity will be introduced  in Sec.~\ref{s:statfluidapprox} for the fluid sector, but  it is not necessary for the gravity sector of our system if the IWM\ approximation is employed.
 
Note that the IWM approximation, while technically convenient, does not allow for purely outgoing gravitational radiation, as would be present in an isolated binary in nature. In particular, the metric of a non-spinning binary system is known to no longer be spatially conformally flat beyond the first post-Newtonian approximation: See, e.g., the discussion in~\cite{JohYunTic09,Schaefer2009} in the quasicircular case, and~\cite{RieSch96} for some work including eccentricity. The assumption of conformal flatness is thus thought to be responsible for at least some of the initial spurious radiation observed at the beginning of all numerical relativity simulations of compact binaries. (One finds reductions in some components of the initial spurious radiation when one drops the assumption of conformal flatness  
in the binary black hole case, e.g.,~\cite{Lov08, MunKelZlo11, ReiTic12,Chu14,ThrYY}.) The \textit{waveless} approach \cite{UryLimFri06,UryLimFri09} involves a (constraint-solved) construction of binary neutron star data that does not  assume spatial conformal flatness, with
some unpublished evolutions~\cite{ShiYY}, but in general this aspect has not been studied nearly as well for binary neutron stars as for binary black holes. Gravitational waves can naturally be accommodated in the \textit{fully constrained formulation} of general relativity \cite{BonGouGra03,CorCerDim09,CorCerIba11,CorCerIba12}.  Nevertheless,  the overall physics of the simulation is not significantly affected by this aspect of the initial data (see~\cite{GarLovKid12} in the binary black hole case), 
though the high-frequency spurious radiation can decrease the accuracy of the simulation~\cite{ZloPonLou12,ZhaSzi13}.

\section{Stationary fluid approximation}
\label{s:statfluidapprox}

\subsection{Circular orbits \&\ helical symmetry}
\label{s:circorbitshelsym}

Binaries on circular orbits  possess a \textit{helical Killing vector}
 which generates time translations in a rotating frame:
\be \label{eq:helicalKV}
k^\alpha = t^ \alpha 
+ \Omega \varphi ^\alpha= t^ \alpha + \Omega \,(x \, y^\alpha - y \,x^\alpha).
\ee
The vectors $\vv{t} = \vv\partial_t$,  $\vv{x} = \vv\partial_x$, and
$\vv{y} = \vv\partial_y$ generate translations in the $t$, $x$, and $y$ directions, respectively, while $\vv{\varphi} = \vv\partial_\varphi$ generates rotations in the $\varphi$ direction (i.e., about the $z$-axis);  $\Omega$ is the orbital frequency.
For circular orbits, the system appears  stationary in a  frame corotating
with the binary.  Hence,  by virtue of Noether's theorem \eqref{eq:injectionEnergyNoether} and \eqref{eq:injectionEnergyBernoulli1}, the
\textit{energy in a rotating frame},
\be \label{eq:injectionEnergy}
\mathcal{E}=-k^\alpha \pi_\alpha
\ee
is conserved along streamlines.
The conserved quantity  $\mathcal{E}$, often called the \textit{injection energy} \cite{FriedmanStergioulas2013}, is analogous to the Jacobi constant of motion of test particles around Newtonian circular binaries \cite{Car79}.
For irrotational or corotating  binaries on circular orbits, this quantity  is constant throughout each star  and is extremely useful for constructing equilibrium models numerically via self-consistent field methods \cite{PriMarFri09,BonGouMar98b,Teu98,Gou98a,FouKidPfe08,Shi98}.  
As mentioned earlier, this conservation law is equivalent to (a relativistic generalization of) the \textit{strong} \textit{Bernoulli principle} and  follows quickly  from  the Cartan identity \cite{Car79,Gou06} applied to any   vector field $k^\alpha$ that Lie-derives the flow:
\be   \label{eq:Cartank}
\Lie{\vv k}  \pi_\alpha =  k^\beta   ({\rm{d}} \pi)_{\beta \alpha} + \nabla_\alpha(k^\beta \pi_\beta)=0.
\ee
For flows where $u^\alpha$ is parallel to $k^\alpha$ (such as rigid rotation or rigid translation),  
 the first term in the above equation vanishes by virtue of the Euler equation  \eqref{e:cold-Euler}. In the more relevant case where  the flow is \textit{irrotational}, i.e., the canonical vorticity vanishes,  we have
\be \label{eq:irrotational}
\pi_\alpha=\nabla_\alpha \Psi  \Leftrightarrow 
({\rm{d}} \pi)_{\alpha \beta}=\nabla_\alpha {\rm{}} \pi_\beta-\nabla_\beta {\rm{}} \pi_\alpha=0,
\ee
for some velocity potential $\Psi$. The Euler equation \eqref{e:cold-Euler}
is thus automatically satisfied and the first term in the identity
\eqref{eq:Cartank} again vanishes, implying that the injection energy \eqref{eq:injectionEnergy} is constant throughout the star:
\be   \label{eq:injencon}
\nabla_\alpha\mathcal{E}=0.
\ee
For many binary neutron stars, the stars' spin frequency is much smaller than the orbital frequency for the last ten to twenty orbits before
merger, making irrotational flow a natural approximation, as is discussed in Sec.~\ref{s:intro}.
  
\subsection{Constant three-velocity approximation}
\label{s:constvapprox}

For any barotropic flow,  Kelvin's circulation theorem guarantees that a flow that is initially irrotational will remain irrotational \cite{FriedmanStergioulas2013,Markakis2011a}. This result is exact for time-dependent 
 spacetimes without symmetries. Hence, in what follows, we shall allow for eccentricity but we will retain the assumption of \textit{irrotational  flow}. To construct irrotational initial data, one typically  substitutes Eq.~\eqref{eq:irrotational} into the continuity equation \eqref{eq:continuityeqn} and numerically solves the resulting equation
\be \label{eq:velpotcont}
\nabla_\alpha \left( \frac{\rho}{h} \nabla^\alpha \Psi \right)=0
\ee
for the velocity potential.  This equation requires boundary conditions on
the star surface and thus surface-fitted coordinates are typically used. For
elliptic solvers based on Cartesian multigrid methods, this is technically
difficult. 

\begin{figure}[t]
  \centering
   \includegraphics[width=0.49\textwidth]{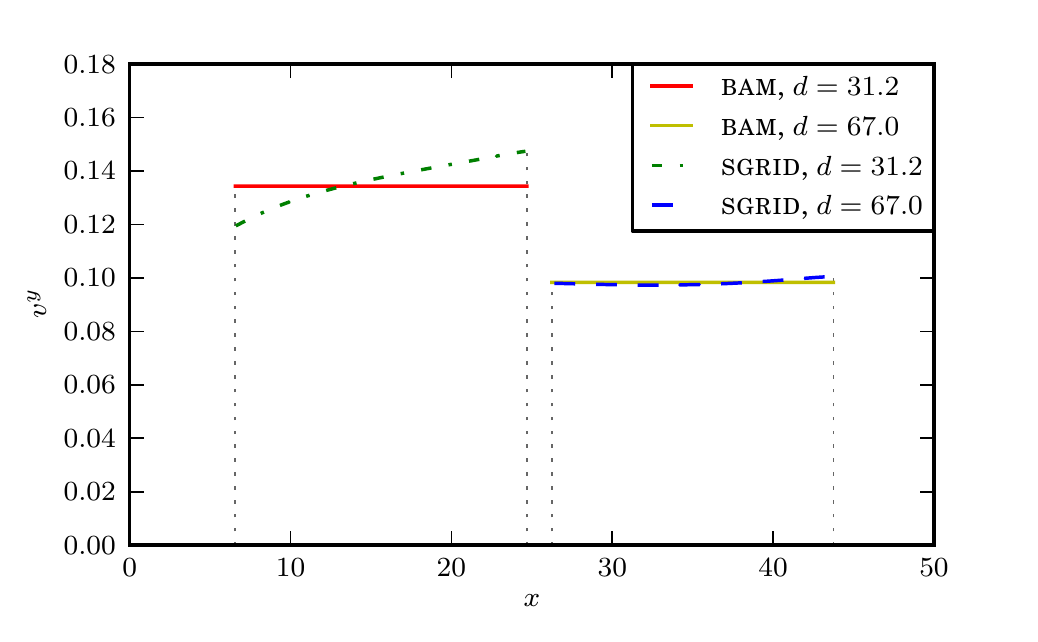}
    \caption{ \label{fig:vy_comparison}Comparing equilibrium data constructed with the constant fluid velocity approximation (using the implementation of our method in {\sc{bam}}) to    data constructed by solving for the  exact velocity potential with {\sc{sgrid}}.     We show 
    the fluid velocity along the $x$-axis for two different values of $d$, the coordinate separation of the stars' centers. One can see a deviation of $\sim 10\%$ for $d=31.2$, but this deviation decreases to $\sim 1\%$ for $d=67$. We have cut the data at the surface of the star, denoted by the vertical thin dotted lines, since the velocity is only well-defined in the star's interior.
    } 
\end{figure}

Nevertheless, multigrid methods are widespread, as they provide a
simple way to test a new method and allow one to solve on the grid used for
evolution, avoiding the need to interpolate as well as the need for
surface-fitted coordinates (which are required for spectral solvers, such as
{\sc{lorene}}~\cite{LORENE} and {\sc{sgrid}}~\cite{Tic09a}). In particular, there is a multigrid method implemented in the \bam code~\cite{BruGonHan06, ThiBerBru11} that we also use for evolutions. For such multigrid implementations, instead of solving for
the velocity potential, it can be convenient to 
approximate the fluid three-velocity measured by  coordinate observers as \textit{homogeneous}, i.e., constant
throughout the fluid.
In particular, if the neutron stars are initially at
apoapsis, with the center of each star located on the $x$-axis,  then we assert that each companion initially moves
rigidly along the $y$-direction with instantaneous 
four-velocity field approximated
by
\be   \label{eq:uboosted}
u^\alpha = u^t  (t^\alpha + v^y\, y^\alpha). 
\ee
Here,  the parameter $v^y:=u^y/u^t=dy/dt$ 
denotes the instantaneous  three-velocity of a fluid element measured by a coordinate observer, which we approximate as constant throughout the star. Note that, in general,  this parameter has  a different sign and magnitude for each star. For  irrotational incompressible flows, this quantity is \textit{exactly} constant throughout the star in the Newtonian  limit (cf.\ the discussion in the Appendix). For irrotational, relativistic, compressible flows,  this assumption is valid in an  approximate sense,  as  illustrated in Fig.~\ref{fig:vy_comparison}.
A direct comparison with exact irrotational initial data (obtained by solving for the velocity potential)  shows that the parameter $v^{y}$ is approximately constant to an accuracy of $\sim 1\%$ when the neutron stars are $67 \simeq 100\text{ km}$ apart (measured by the coordinate separation of their centers; recall that we take $M_\odot = 1$). This accuracy degrades to $\sim 10\%$ when the separation decreases to $31.2 \simeq 47 \text{ km}$.
The Lorentz factor in the above equation is  determined from the normalization condition $g_{\alpha \beta}u^\alpha u ^\beta=-1$, which yields
\be\label{eq:ut}
{u^t} = {[ - {g_{tt}} - 2{g_{ty}}v^y - {g_{yy}}{(v^y)^2}]^{ - 1/2}}.
\ee
The Lorentz factor $u^t$ is not assumed to be constant.

We note that the constant three-velocity approximation \eqref{eq:uboosted}  is \textit{optional}.
It merely provides a way of easily constructing   initial data for  approximately irrotational stars, on circular or eccentric orbits,
using multigrid elliptic solvers. One can still opt to solve Eq.~\eqref{eq:velpotcont}   
for the exact velocity potential with a solver that uses surface-fitted coordinates, using a method outlined at the end of the next subsection.

\subsection{Eccentric orbits \&\ instantaneous helical symmetry}
\label{s:eccorbitshelsym}

Binaries on eccentric orbits  are not  stationary in a rotating frame and  thus lack helical symmetry. Hence, no self-consistent method  for constructing initial data existed  for eccentric binaries to date, and numerical relativity groups have resorted to using a superposition of boosted spherical stars as initial data, possibly solving the constraints, but not solving the Euler equation. 
Use of such initial data entails not one, but three  distinct physical approximations regarding the fluid configuration: (i) that the stars are spherical and thus lack tidal deformation, (ii) that the fluid is stationary in a linearly comoving frame, and (iii) that the initial velocity field is homogeneous, i.e., constant throughout the fluid.
The above approximations are only valid at infinitely large separation and are violated as the stars inspiral towards each other. In addition, the metric construction as a linear superposition of boosted spherical star data entails (iv) violations of the Einstein constraint equations.

Thus, it remains an open question whether the $f$-mode oscillations observed in  simulations of highly eccentric binary neutron stars~\cite{GolBerThi11,EasPre12} and black hole-neutron star binaries~\cite{SteEasPre11,EasPreSte12} are spuriously excited due to inconsistent initial data.
In what follows, it will be demonstrated that the above  approximations  are  distinct and may in fact  be relaxed one by one.
This allows one to examine whether removing certain approximations  removes spurious oscillations  in the early part of  inspiral simulations.
In particular, it will be shown that dropping assumptions (i),
 (iv) and retaining assumptions (ii), (iii), leads to oscillations of the same magnitude as for boosted spherical stars. On the other hand, the oscillations are nearly eliminated if assumption (iii) is retained but assumption (ii) is dropped. That is, \textit{stationarity in a linearly comoving frame is the main source of error} in boosted spherical star  initial data. In the circular limit, the data should  be \textit{stationary in a rotating frame}. One can quantify the discrepancy between these two symmetries by testing the former symmetry against exact circular initial data stationary in a rotating frame; the discrepancy (10--20\%) is comparable in magnitude to the  central density  oscillations in simulations.
Thus, for eccentric binaries,
the main difficulty is  to define a suitable notion of stationarity, i.e., to obtain a generalization of  the vector field  \eqref{eq:helicalKV}
for which the   energy \eqref{eq:injectionEnergy} is approximately constant.

 We shall assume that  the neutron stars are initially at
apoapsis, with the center (i.e., point of maximum density) of the star
of mass $m_1$, $m_2$ located respectively at position
\begin{subequations}
\label{eq:centersx1x2}
\bea
x_1&=& a(1+e)\frac{m_2}{m_1+m_2} + x_\text{cm},\,\ \\
x_2&=&-a(1+e)\frac{m_1}{m_1+m_2} + x_\text{cm}
\eea
\end{subequations}
on the $x$-axis, where $x_\text{cm}$ denotes the initial position of the center of mass (on the $x$-axis).
Here, $a$, $b$, and $e=(1-b^2/a^2)^{1/2}$ denote the semimajor axis, 
semiminor axis, and eccentricity\footnote{The parameter $e$ estimated via this  formula is only used as input to monotonically control    the  eccentricity of the orbit obtained upon evolving the initial data. Due to finite-size and relativistic effects,  the actual orbits deviate from closed ellipses and the  orbital eccentricity deviates from the Newtonian point particle limit, cf.\ Secs.~\ref{s:radvel} and~\ref{ssec:trackswaves}.}  of the ellipse traced by the vector 
joining the two star centers and the binary's center of mass.
We have selected the  \textit{apoapsis} for our construction of initial data because this represents a moment of \textit{time symmetry} of the radial motion, in the sense that  the radial velocity vanishes.  In addition, maximizing the distance between the two stars  happens  to maximize the accuracy of our approximations.

We seek  a   vector field $k^\alpha$  that approximately Lie-derives the flow. If such a vector exists, and the flow is irrotational, then the Cartan identity  \eqref{eq:Cartank} will give an injection energy of the form  \eqref{eq:injectionEnergy} that is approximately constant throughout the fluid. In light of Eq.~\eqref{eq:uboosted} and the fact that the difference between a circular orbit and an eccentric orbit  lies in the magnitude of the $y$-component of the initial 3-velocity at the apoapsis, our \textit{ansatz} consists of  generalizing the helical vector field \eqref{eq:helicalKV} by adding a boost along the $y$-direction.
This amounts to a change in  the  center of rotation of the helical vector, and we shall use both viewpoints interchangeably in what follows. These considerations lead us to introduce what we term an \textit{instantaneously} \emph{inscribed helical vector} (or, more informally, a \emph{``helliptical'' vector}), 
\be \label{eq:hellipticalKV}
k^\alpha = t^ \alpha 
+ \omega \, \varphi ^\alpha + \lambda \, y^\alpha= t^ \alpha + \omega \, [(x-x_{\rm c}) y^\alpha - y \,x^\alpha],
\ee
with a   rotation frequency $\omega$  that now differs from the orbital frequency. The  displacement $x_{\rm c}$, or boost parameter   $\lambda = - \omega \,x_{\rm c}$,  will be determined on physical grounds (with $\lambda$ and $x_{\rm c}$ different for each star).
The  choice  $\omega=0$ and $\lambda=v^{y}$ would lead to $k^\alpha=t^\alpha + v^{y}\, y^\alpha$, implying stationarity in a linearly comoving frame. 
As mentioned earlier, constructing initial data with this assumption yields spurious oscillations of similar magnitude to  boosted spherical stars and thus   explains why assumption (ii) described above must be dropped.\footnote{In particular, this choice would mean that $(\partial_t + v ^{y}\partial_y)h$ vanishes. For circular orbits, one can check this against the exact enthalpy profile, which  satisfies   $(\partial_t + \Omega \partial_\varphi)h=0$. Using this exact relation to eliminate the time derivative, the previous expression becomes $(\partial_t + v^{y}_{1,2} \partial_y)h=\Omega[(x_{1,2}-x)\partial_y+y\partial_x]h$ for star $1,2$. 
This quantity does not vanish in the circular limit  except at the stellar center $x=x_{1,2}$, $y = 0$. 
In addition,  the force equation [of the form \eqref{eq:injectionenergyspatial}] one obtains from the injection energy  \eqref{eq:injectionEnergy} that is constructed from $k^\alpha = t^\alpha + v^y\, y^\alpha$ violates true force balance at the center of the star.  The  violation stems from  the absence of  centrifugal forces, which leave  gravitational forces unbalanced. This results in density oscillations of order $10-20\%$ in simulations,  as illustrated in Fig.~\ref{fig:densosc}.} Instead, for eccentric binaries,  the parameter $\omega$ should be nonzero and  $\lambda$ (or $x_c$) should vanish in the circular limit.

 The  $\omega$ parameter can be determined by requiring that Eq.~\eqref{eq:Cartank}
holds exactly for incompressible binaries on Newtonian eccentric orbits  (cf.\  the discussion in the Appendix).  This yields
\be \label{eq:vomegae}
v^{y}_{1,2} = (1-e)\, \omega \, (x_{1,2} - x_\text{cm}).
\ee
The positions of the star centers, $x_{1,2}$, are given by  \eqref{eq:centersx1x2}
and 
\be  \label{eq:omegaeOmega}
\omega= (1 + e)^{-1}{(1 - {e^2})^{-1/2}}  \bar\Omega,
\ee
where $\bar\Omega=2 \pi /T$ is the mean motion and $T$ is the orbital period.
The $\lambda$ or $x_c$ parameter can then be determined by requiring  $k^\alpha$ to be initially parallel to $u^\alpha$ at the star center. Substituting \eqref{eq:centersx1x2} into \eqref{eq:hellipticalKV} and comparing to \eqref{eq:uboosted} yields
\be \label{eq:lambdave}
\lambda_{1,2}  = - \omega \,x_{\rm c_{1,2}}=  - v^{y}_{1,2}\frac{e}{{1 - e}} - \omega\,x_{\rm{cm}},
\ee
with $v^{y}=v^{y}_{1,2}$ given by \eqref{eq:vomegae}.
As expected, for $e=0$ the inscribed helical symmetry vectors~\eqref{eq:hellipticalKV} yield initial data for circular orbits, while $e=1$ corresponds to zero tangential velocity $v^y_{1,2} = 0$ for a fixed $\omega$, giving a head-on collision starting from rest. [Note that $\omega$ blows up as $e\nearrow 1$. This can be seen in Eq.~\eqref{eq:vomegae}, which our method reproduces in the Newtonian limit, as discussed below, even though we do not use that equation to determine $\omega$. However, one can simply set $\omega = 0$ and $v^y_{1,2} = 0$ in our method to obtain an exactly head-on collision.]

The parameters $\omega$ and $\lambda$ in the
inscribed helical symmetry vectors in Eq.~\eqref{eq:hellipticalKV} have been
determined such that we get the correct answer for irrotational
incompressible stars moving on ellipses 
(see the Appendix).

There is, however, a completely different way to obtain this result.
All we need are the following three reasonable assumptions:
(i) There exists  a vector $k^\alpha$ that approximately Lie-derives the flow. 
(ii) $k^\alpha$ is along the motion of the star center.
(iii) Each star center moves along a segment of an elliptic orbit at apoapsis.

\begin{figure}[t]
  \centering
    \includegraphics[width=0.49\textwidth]{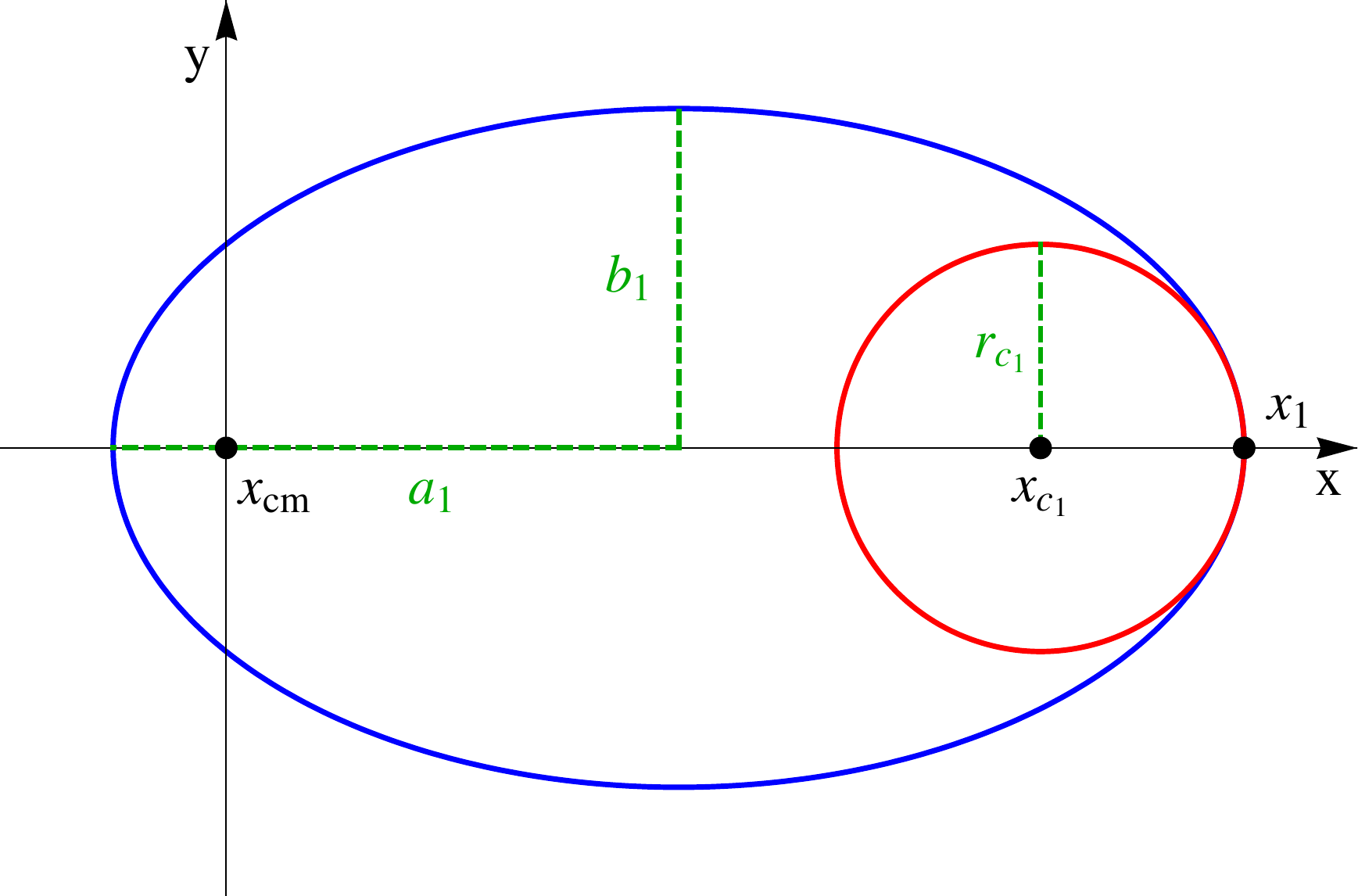}
    \caption{ \label{fig:situation}
Illustration of the approximation of the orbits by using circles inscribed
into the orbital ellipse in a way that their curvature is the same as the
one of the ellipse. We show the scaled semimajor axis
$a_1=a\frac{m_2}{M}$, semiminor axis $b_{1}=b\frac{m_2}{M}$, and the radius
$r_{{\rm c}_{1}} =b_1^2/a_1$ and center $x_{{\rm c}_{1}}$ of the inscribed circle
as well as the center $x_1$ of one star. The center of mass is denoted by
$x_\text{cm}$.}
\end{figure}

Assumption (ii) is absolutely necessary, otherwise $k^a$ can never be an
approximate Killing vector.  Assumption (iii) specifies what orbit we want. It
seems reasonable that at least approximately we should have Newtonian and
thus elliptic orbits. Since we only need a small segment of an orbit near apoapsis, we will
approximate this segment by the circle inscribed into the elliptical orbit there
(see Fig.~\ref{fig:situation}), i.e., the
circle that that has the same curvature radius $R_{\rm{c}}$ as the ellipse at
apoapsis. From elementary considerations it is clear that $R_{\rm{c}}$ has to be
$R_{\rm c}=(1-e^2)A$ for an ellipse with semimajor axis $A$ and eccentricity $e$.
In Newtonian theory it is well known that two particles of masses $m_1$ and
$m_2$ that orbit around each other, move on ellipses with semimajor axes
$a_1 = \frac{d_1}{1+e}$ and $a_2 = \frac{d_2}{1+e}$,
where $d_1$ and $d_2$ are the distances of the particles from the center of
mass at apoapsis. Thus the radii of the inscribed circles have to be
\be
\label{eq:r_c_of_d}
r_{{\rm c}_{1,2}}  = (1-e) d_{1,2} .
\ee

Obviously these two inscribed circles are not centered on the center of mass,
but on the points
\be
\label{eq:x_circlecenter}
x_{{\rm c}_{1,2}}= x_{1,2} \mp r_{{\rm c}_{1,2}}= x_\text{cm} + e (x_{1,2} - x_\text{cm}) ,
\ee
where we have used $d_{1,2} = |x_{1,2} - x_\text{cm}|$ and assumed that apoapis
occurs on the $x$-axis. (The upper and lower signs
correspond to the subscripts $1$ and $2$, respectively.)
Assumption (ii) then tells us that the approximate Killing vector must have
the form
\be 
\label{eq:ellKV}
k^\alpha_{1,2} = 
t^ \alpha  + \omega_{1,2} \, [(x-x_{{\rm c}_{1,2}}) y^\alpha - y \,x^\alpha]
\ee
near each star. While the expressions for $k^\alpha_{1}$ and $k^\alpha_{2}$
in Eq.~(\ref{eq:ellKV}) look different, we will show next that 
$\omega_{1}=\omega_{2}$, so that far from the stars (where $x\gg x_{\text{c}_{1,2}}$) there is only one
approximate Killing vector. From the definition of the center of mass, we obtain
$m_1 d_1 = m_2 d_2$ and thus using Eq.~(\ref{eq:r_c_of_d}) we find
\be
\label{eq:mr_rel}
m_1 r_{{\rm c}_{1}}= m_2 r_{{\rm c}_{2}}.
\ee
If we assume that the center of mass is at rest,
momentum conservation demands that
$m_1 \omega_{1} r_{{\rm c}_{1}} = m_2 \omega_{2} r_{{\rm c}_{2}}$. Together with
Eq.~(\ref{eq:mr_rel}), this implies that
\be
\omega_{1}=\omega_{2}=:\omega .
\ee
Finally, in order to satisfy assumption (ii), $v^y$ in
Eq.~(\ref{eq:uboosted}) must be chosen to be
\be
\label{eq:v1_2}
v^{y}_{1,2} = \pm \omega r_{{\rm c}_{1,2}} = (1-e) \omega (x_{1,2}-x_\text{cm}) .
\ee

The value of $\omega$ is usually derived from a ``force balance" equation, which has the form of 
Eq.~\eqref{eq:injencon} applied at the center of each star.
In the Newtonian limit, this equation reads
\be \label{eq:injectionenergyspatial}
\bar D_i \mathcal{E}=0,
\ee
where the injection energy in star 1 is given by 
\be
\mathcal{E} =
 \frac{1}{2}v^2 + \Phi+h  - \omega \, [(x-x_{{\rm c}_{1}})\, v^y - y \, v^x] 
\ee
and $\Phi$ is the Newtonian gravitational potential. Here $\bar{D}_a$ is the covariant derivative compatible with the Euclidian 3-metric $f_{ab}$ in $\mathbb{E}^3$. (We have used the same
notation for this covariant derivative as for the covariant derivative compatible with the conformal $3$-metric, since we took the conformal $3$-metric to be flat.) If the stars are far apart and thus almost spherical, then the orbits are almost elliptic and we can approximate the potential due to star 2 as that of a point mass, given by $\Phi=-G m_2/\|\vv{x} - \vv{x}_2\|$, and neglect the gradient of star 1's potential at its center. (We show factors of $G$ explicitly here and during the rest of this discussion to make the distinction between the gravitational and centripetal forces immediately apparent.)
If we use this expression in Eq.~(\ref{eq:injectionenergyspatial}) at the
star center $x=x_{1}$ where the enthalpy is maximum, $\partial_x h = 0$, we find
\be
G \frac{m_2}{r^2} - \omega v^y = 0,
\ee
where $r$ is the separation of the stars.
Using $v^{y} = v^y_1$ from Eq.~(\ref{eq:v1_2}) we find 
\be
G \frac{m_1 m_2}{r^2} = m_1 \omega^2 r_{{\rm c}_{1}},
\ee
which implies that $\omega$ is such that the centripetal force needed to
keep star 1 on the inscribed circular orbit is provided
by the gravitational force due to star 2.

Thus the ansatz~\eqref{eq:hellipticalKV} admits a beautifully simple geometrical
interpretation. The vector $k^\alpha$ can be interpreted as an
\textit{instantaneously inscribed helical vector field}, that generates time
translations in a frame rotating about a point 
$(x_{{\rm c}_{1,2}},0,0)$
given by Eq.~(\ref{eq:x_circlecenter}).
When projected onto a  spatial slice $\Sigma_t$, the integral curves of this vector field are
\textit{circles inscribed into the elliptic trajectory at the apoapsis}, with
their center at $(x_{{\rm c}_{1,2}},0,0)$ and a radius $r_{{\rm
c}_{1,2}}$ equal to the
\textit{radius of curvature of the ellipse} at the apoapsis. Contrary to the
circular case, the integral curves of this vector field do not coincide with
the (eccentric) orbit of the star center (though they are tangent to each other at $t=0$, which suffices for constructing initial data),  and the energy
\eqref{eq:injectionEnergy} is not conserved by the flow. 
Instead, as one may see by operating with $k^\alpha \partial_\alpha$ on (A16), this
energy is  constant spatially [Eq.~\eqref{eq:injectionenergyspatial} is satisfied throughout the fluid] at the initial time
$t=0$ but not for $t>0$.
 Because the spatial derivatives of the injection energy $\mathcal{E}$ approximately vanish initially, we infer that 
its time derivative also vanishes instantaneously
 (i.e. $\partial_t \mathcal{E}=-k_{1,2}^a \partial_a \mathcal{E}=0 $ at $t=0$).
 Eq.~\eqref{eq:injectionEnergyBernoulli1}
is then satisfied and, in this sense, $k_{1,2}^\alpha$ may be considered  approximate Killing vectors for the region of spacetime occupied by each star.
 
 In principle, if one wishes to construct initial data at any time (not necessarily at apoapsis), one could use a more general ansatz,
$k^\alpha = t^ \alpha + \omega \, [(x-x_{\rm c}) y^\alpha - (y-y_{\rm{c}}) x^\alpha]$,
with  the parameters $\omega$, $x_{\rm c}$ and $y_{\rm c}$ determined by assuming that the orbit is \textit{instantaneously circular}, i.e., by constructing a circle inscribed at the elliptic trajectory at the point of interest.
We have not tested this more general construction, as nonvanishing radial derivatives and close separation are expected to degrade accuracy. Instead, for reasons detailed above, we  opt to construct initial data at apoapsis and set $y_{\rm{c}}=0$. Note that the stars will likely obtain a small initial radial velocity component when the data are evolved, from the ``kick'' due to the initial spurious radiation. However, at the relatively large separations we are considering, this radial velocity component will be small, so the stars' orbit will still begin very close to periapsis, thus retaining all of the favorable features which led us to choose that point.

With the ansatz  \eqref{eq:hellipticalKV}, the first integral  \eqref{eq:injectionEnergy} to the relativistic Euler equation \eqref{e:cold-Euler}
becomes
\be \label{eq:injectionenergyeccentric}
\mathcal{E}=-h\{u_t + \omega \, [(x-x_{\rm c})\, u_y - y \, u_x]\},
\ee
where
\be \label{eq:usubstition}
u_\mu = g_{\mu \nu} u^\nu= u^t(g_{\mu t}+v^{y} \, g_{\mu y})
\ee
if the constant 3-velocity approximation  \eqref{eq:uboosted}
is used, or 
\be
u_\mu= h^{-1} \nabla_\mu \Psi
\ee
if one solves Eq.~\eqref{eq:velpotcont} for the exact velocity potential.
In this case, time derivatives are eliminated in favor of spatial derivatives using the replacements 
$\partial_t \Psi \rightarrow -\mathcal{E}\ -k^a D_a \Psi$,
$\partial_t \rho \rightarrow -k^a D_a \rho$ 
[with $k^a$ denoting the spatial part of the vector field \eqref{eq:hellipticalKV} and $D_a$ the covariant derivative
compatible with the physical $3$-metric $\gamma_{ab}$], resulting in the equation 
\cite{FriedmanStergioulas2013}
\be \label{eq:velpotcont2}
D_a D^a  \Psi = (\beta^a+k^a) D_a \Lambda -[D^a \Psi-\Lambda (\beta^a +k^a)]D_a \ln \frac{\alpha \rho}{h},
\ee
where $\Lambda:=\alpha^{-2}[\mathcal{E}+(\beta^a+k^a)D_a \Psi]$. For fixed $h$ and $\rho(h)$, this equation is elliptic [the principal part is $\gamma^{ab} + \ell^a\ell^b$, $\ell^a := \alpha^{-1}(\beta^a + k^a)$, which is clearly positive definite, since the $3$-metric is positive definite]. It can be solved iteratively  for $\Psi$ using a numerical Poisson solver, with the right hand side (including the additional terms one obtains if one does not assume spatial conformal flatness) treated as a fixed source in each iteration, as in~\cite{UryLimFri09}. The   boundary condition 
\be \label{eq:velpotcont2bound}
[D^a  \Psi - \Lambda(\beta^a+k^a)]D_a \rho |_{\rm S}=0
\ee
is imposed on the surface of the star \cite{FriedmanStergioulas2013}.

 Taking the Newtonian limit of the energy \eqref{eq:injectionenergyeccentric} (cf.\ the discussion in the Appendix) and applying Eq.~\eqref{eq:injectionenergyspatial} at the center of one star [given by \eqref{eq:centersx1x2}
and defined as the point of maximum specific enthalpy, $\partial_x h|_{x=x_{1,2}}=0$]
yields a force balance equation $\partial_x \mathcal{E}|_{x=x_{1,2}}=0$. It is  reassuring and straightforward to check that, for inverse square  gravitational forces, this equation amounts  to Kepler's third law for eccentric binaries
\be \label{eq:Kepler3rd}
\bar\Omega^2=\frac{G(m_1+m_2)}{a^3},
\ee
where $a$ is the  semimajor axis of the ellipse traced by the vector joining the two star centers.

\subsection{Radiation reaction \& radial velocity}
\label{s:radvel}

When the eccentricity parameter $e$ is set to zero, the ansatz~\eqref{eq:hellipticalKV} reduces to the helically symmetric ansatz~\eqref{eq:helicalKV}. However, because this expression neglects the radial velocity due to radiation reaction,  responsible for  binary inspiral,
evolutions of helically symmetric data
exhibit residual eccentricity, e.g., the orbital
separation acquires an oscillatory contribution and does not decrease monotonically in time~\cite{KyuShiTan14}. In light of the preceding discussion, a way to incorporate radial velocity is to  include a (constant  as measured by coordinate observers) three-velocity $v^x$ along the radial (initially $x$-) direction in $u^\alpha$ and $k^\alpha$, thereby replacing Eqs.~\eqref{eq:uboosted} and \eqref{eq:hellipticalKV}
by
the ansatz
\be   \label{eq:uboostedR}
u^\alpha = u^t  (t^\alpha +v^x\, x^\alpha+ v^y\, y^\alpha) 
\ee
and
\be  \label{eq:hellipticalKV2}
{k^\alpha } = {t^\alpha } +\omega( x-x_{\rm{c}}) {y^\alpha } - (\omega y - {v^x}){x^\alpha}.  
\ee

The relation between $\omega x_{\rm{c}}$ and $v^y$ is again given by Eq.~\eqref{eq:vomegae}, which guarantees that the condition $u^\alpha=u^t k^\alpha$ is satisfied at the center $x=x_{1,2}$ of each star.

 The radial velocity parameter $v^x$ can be determined, for example, by post-Newtonian or effective-one-body theory, with or without tidal (finite-size) corrections for the given equation of state. This has been sufficient to significantly decrease the eccentricity in simulations in the past (for example, the  simulations performed in \cite{ReaMarShi09}  used a Lorentz boost of the initial data based on point-particle post-Newtonian values for the radial velocity and led to significantly lower eccentricity than the simulations in \cite{ReaBaiCre13} that did not incorporate radial velocity). 
 Alternatively, if one wishes to obtain truly quasicircular data, then one
may set $e=0$ and an initial value for $v^x$ in the above equations, evolve
the resulting data for a period of time adequate to determine the orbital
eccentricity, adjust $v^x$ to reduce the eccentricity, and iterate this process until the resulting eccentricity is
sufficiently small, similarly to what has been done for binary black hole
simulations \cite{PfeBroKid07,HusHanGon07,TicMar10,MroPfe12,
PueHusHan12}.\footnote{While the present manuscript was near completion, we
were informed that recently this procedure was also implemented in
\cite{KyuShiTan14} for binary neutron star simulations.}

As stated earlier, the approximation~\eqref{eq:uboostedR} is meant to be used when elliptic solvers without surface-fitted coordinates are employed.  If surface-fitted solvers are available, one may instead solve the elliptic equation \eqref{eq:velpotcont2} for the velocity potential, with $k^\alpha$ given by \eqref{eq:hellipticalKV2}. In either case, the approximate first integral to the Euler equation is given by \eqref{eq:injectionEnergy}, and provides the basis for a self-consistent iteration that will be outlined below.

If the stars are represented by compact monopole sources that inspiral towards their center of mass with a time-dependent radial velocity, then one can straightforwardly show that the vector  \eqref{eq:hellipticalKV2}  Lie-derives the Newtonian gravitational field in a zone near   each monopole. In this sense, the vector field is  approximately Killing  near each compact star. Note that this is not true  globally and that the vector field is different for each star;  this does not limit our formulation since $k^\alpha$  is merely used for  the hydrodynamics inside each star and not for the gravitational field equations.

\section{Numerical method}
\label{s:num_meth}
\subsection{Elliptic solver}
To construct binary neutron star initial data (i.e., expressions for the five elliptic quantities $\psi$, $\alpha$, and $\beta^i$ and the matter density profile, in our constant $3$-velocity approximation), we solve the five elliptic equations \eqref{eq:cts2} together with the first integral of the Euler equation \eqref{eq:injectionenergyeccentric}, where the latter has to be fulfilled throughout the stars. We solve these equations by iteration using a self-consistent field method. In each step we approximate derivatives by standard second-order finite differencing operators in a full approximation storage (FAS) multigrid scheme  with nested boxes on a Cartesian grid. We employ a red-black Gau\ss-Seidel method \cite{PreFlaTeu07a} for our relaxation scheme. This method should lead to second-order convergence, since it uses second-order finite differencing. However, if desired, the computation could be extended to up to eighth-order finite differencing using the \textsc{olliptic} code \cite{GalBruCao10}, and 
therefore yield higher convergence orders. Another way to increase the accuracy and obtain spectral convergence would be to implement the method in {\sc{sgrid}}.

The iteration process is highly sensitive to the initial guess and cannot be started with an arbitrary set of values. Solutions to Einstein's equations for isolated non-rotating neutron stars are well known and will serve as an initial guess for our iteration. We proceed as in \cite{GolBerThi11}. We first construct two single relativistic non-rotating spherical stars (solutions to the Tolman-Oppenheimer-Volkoff [TOV] equations \cite{Tol39,OppVol39}) with the same baryonic mass or central density (depending on what we fix during the iteration) as that desired for the stars in the initial data. We then boost these stars with a Lorentz transformation in the $\pm y$ direction to give the appropriate orbital motion and superpose the resulting $4$-metrics by
\begin{equation}
\label{e:sup}
g_{\mu\nu}^{\text{(sup)}}=g_{\mu\nu}^{(1)}+g_{\mu\nu}^{(2)}-\eta_{\mu\nu},
\end{equation}
where $g_{\mu\nu}^{(A)}$ denotes the metric of star $A\in\{1,2\}$ (including the boost) and $\eta_{\mu\nu}$ is the Minkowski metric. We extract initial values for the elliptic quantities from the superposed $4$-metric and also initialize the matter enthalpy profile using the TOV solution. Since the spatial metric is not conformally flat, due to the boost, we simply take $\psi^4$ to be the $xx$ component of the $3$-metric. Note that one can usually take the boost to be zero without affecting the convergence if the stars are widely separated (as in all the eccentric runs we show in Sec.~\ref{s:ecc}). However, a nonzero boost is necessary to obtain convergence when the stars are close (as in the $d = 31.2$ run shown in Fig.~\ref{fig:vy_comparison}); here we set the boost parameter by hand to reproduce the expected fluid velocity.

The grid is simple Cartesian, and hence not compactified in any spatial direction. We choose large grid setups and thus distant outer boundaries, which are located far away from the stars at a distance of $\sim 3000$, while a typical stellar radius is $\sim 10$. We set Dirichlet boundary conditions for the elliptic variables ($\psi$, $\alpha$, and $\beta^i$) at our outer boundary by using values from the superposed TOV metric there, which we find produces better results (e.g., better agreement with \sgrid at the boundary in the quasicircular case) than just using the values these variables would have at infinity (i.e., $\alpha=\psi=1$ and $\beta^i=0$).

\subsection{Iteration scheme}

In each iteration step we have to compute four constants from the integrated Euler equation \eqref{eq:injectionenergyeccentric} and its derivative with respect to $x$, which---evaluated at the star's center---yields the force-balance equation
\begin{equation}
 0 =\left.\left\{\partial_x u_t + \omega \, \left[(x-x_{\rm c})\, \partial_x u_y + u_y - y \, \partial_x u_{x}\right]\right\}\right|_{x=x_{1,2}}. \label{eq:forcebalance}
\end{equation}
The constants of interest are the orbital frequency $\omega$, the center of mass $x_\text{cm}$, and the injection energy of each star $\mathcal{E}_{1,2}$, which is given as the constant of integration. We are free to make arbitrary choices for the central density and the separation of the stars in advance and fix them throughout the iteration. Note that it is also possible to fix the total rest-mass instead of the central density, which we do when computing sequences. We evaluate both the first integral to the Euler equation \eqref{eq:injectionenergyeccentric} and the force balance-equation \eqref{eq:forcebalance} at the centers of the two stars, which are located at fixed positions $x_{1,2}$.  For unequal mass stars, we have to use a root finder inside our overall iterative scheme to obtain these constants, but for equal mass stars this system is degenerate and $x_\text{cm}$ can simply be set to zero, allowing us to use algebraic solutions without an additional call to a root finder (though we still solve by iteration overall).

Using \eqref{eq:usubstition}, the first 
integral can be written as 
\begin{equation}
\begin{split}
 \mathcal{E}_{1,2} &= - h u^t\{-\alpha^2+\psi^4[\beta^i\beta_i+v_{1,2}\beta^y-\omega \,y\,\beta^x\\
 &\quad +\omega (x - e (x_{1,2}-x_{\rm{cm}})-x_{\rm{cm}})(\beta^y+v_{1,2})]\}
 \label{eq:integratedeuler_equalmass}
\end{split}
\end{equation}
and the force-balance equation can be rewritten in the same way. The latter can be solved for $\omega$ algebraically if we set $h'(x_{1})=0$, which ensures that the maximum density stays at the center of the star. (While the force-balance equation contains $u_t$, which depends on $\omega$, we solve it with $u_t$ fixed, and then update $u_t$ later in the iteration.) The frequency $\omega$ can now be substituted into (\ref{eq:injectionenergyeccentric}), along with value of the enthalpy $h$ obtained from the fixed central density, to obtain the injection energy $\mathcal{E}$.
Using these values for $\omega$ and $\mathcal{E}$, we are able to compute the enthalpy density profile by solving \eqref{eq:integratedeuler_equalmass} for $h$. We then use the equation of state to obtain the mass density $\rho(h)$. We finally update $u_t$ using \eqref{eq:ut} [and \eqref{eq:usubstition} to lower the index] and the new values of the constants and solve the force-balance equation for $\omega$ again with the new $u_t$, iterating over these steps until the change in $\omega$ falls below numerical accuracy, which usually happens after a few iterations.

Having gathered all the necessary constants, we can compute the source terms using \eqref{eq:sourceterms} and then use the multigrid scheme to solve the elliptic equations. In this step, we employ softening, i.e., instead of taking the full value of the updated variable, we use a weighted average of the old and new variables. Specifically, all elliptic variables $X$ are set using $X=0.25X_\text{new}+0.75X_\text{old}$, similar to the softening used in \cite{Tic12}. 
We then return to the computation of the constants, which are no longer valid, since the elliptic variables have now changed. This process is iteratively repeated until the change in the elliptic variables falls below a prescribed threshold.   

\section{ quasicircular ORBITS}
\label{s:quasicirc}

Here we give examples of quasicircular initial data constructed with our method, and verify the code's results against those obtained using {\sc{sgrid}}. We summarize the properties of the initial data sets we 
consider in this section and the following one in Table~\ref{t:idpar}, which also gives the labels we use for the different sets.

We compare the $e = 0$ limit of our data with helical
Killing vector initial data constructed using the spectral code
{\sc{sgrid}}~\cite{Tic09a}, which solves for the velocity potential using
surface-fitted coordinates and also compactifies the grid to include spatial
infinity. For the purposes of our comparison, we use 
$n_A=n_B=24$, $n_{\phi}=8$, $n_c=16$
points in \sgrid (see \cite{Tic09a} for details about {\sc{sgrid}}'s grid
structure), which is sufficiently accurate, due to the code's spectral
convergence.

We perform evolutions of the data using the \bam code, which is a finite difference adaptive mesh refinement code for evolving the Einstein equations~\cite{BruGonHan06}, and includes a high resolution shock-capturing module to solve the equations of relativistic hydrodynamics~\cite{ThiBerBru11}. Specifically, we use the same evolution setup as in~\cite{BerThiBru11}, with the following differences: We use second-order spatial finite differencing for the geometry, consistent with the order of the multigrid algorithm, and fourth-order Runge-Kutta integration in time, along with fourth-order Kreiss-Oliger dissipation (with a factor of $0.5$), as appropriate for second-order spatial finite differencing.
For the evolution of the fluid quantities, we use the (formally) fifth-order
weighted-essentially-non-oscillatory WENOZ scheme found to improve accuracy
in~\cite{BerNagThi12}. Finally, in the gauge conditions, the
coefficient of the contracted Christoffel symbol in the $1+\log$ shift [in Eq.~(15) in~\cite{ThiBerBru11}] has
been set to $1$ instead of the value of $3/4$ used in~\cite{ThiBerBru11,BerThiBru11,BerNagThi12}. (This is a minor change that was made for convenience in~\cite{HilBerThi12} and is not expected to affect anything significantly.)

\begin{table*}[ht]
\caption{\label{t:idpar} Parameters for the initial data sets considered in this paper. Here $m_{1,2}$ denotes the baryonic mass of one of the stars (recall that we are only considering the equal-mass case in this paper), $d$ denotes the initial
coordinate separation of the stars' centers,
$e$ is the
eccentricity parameter set in the initial data, and $\lambda$ denotes the boost parameter used in the inscribed helical symmetry vectors. Additionally, $\kappa$ is the scale parameter in the polytropic EOS, $(\Delta x)_\text{min}$ denotes the finest grid spacing, and ``points'' denotes
the number of points used in each direction on each of $l_{\rm{max}}$ refinement levels (as well as the fundamental grid level $l = 0$, giving $l_\text{max} + 1$ levels total); the levels with $l \geq l_\text{mv}$ are moving. (We do not give $l_\text{mv}$ for the sequence data sets \emph{seq0}--\emph{seq0.9}, which we do not evolve.) 
We name the sets using their eccentricity, with markers for the cases with a different choice of the boost parameter (``v''), data used for a sequence (``seq''), and an additional high-resolution case (``high'').
}
\begin{tabular}{lccccccccc}
\hline\hline
Name      & $m_{1,2}$   &  $d$   & $e$     & $\lambda$ & $\kappa$ & $(\Delta x)_\text{min}$ & points & $l_{\rm{max}}$ & $l_\text{mv}$\\
\hline                                                 
\multirow{2}{*}{ecc$0$}    & \multirow{2}{*}{1.620} & \multirow{2}{*}{$31.3$} & \multirow{2}{*}{$0$}                        &  \multirow{2}{*}{$-\omega x_c$}         & \multirow{2}{*}{$123.65$}       & 0.09375, 0.1875, & 194, 98, & \multirow{2}{*}{5} & \multirow{2}{*}{1}\\
    &   &   &   &  &   &  0.375 & 50 & &\\
ecc$0$v   & 1.620 & $31.3$ & $0$                        &  $v^y$                  & 123.65        & 0.1875 &98 & 5 &1\\
seq$e$  & 1.625 &$[30.64,...,51.44]$    &$0, 0.2, 0.5, 0.9$     & $-\omega x_c $   & 123.65 & 0.156                                &146             &       8       & $\cdots$  \\  
\multirow{2}{*}{ecc$e$}  & \multirow{2}{*}{1.504} & \multirow{2}{*}{$80.0$} & $0.45,0.5, 0.6, 0.73,$      &  \multirow{2}{*}{$-\omega x_c$}         & \multirow{2}{*}{100}      & \multirow{2}{*}{0.25}     & \multirow{2}{*}{130} & \multirow{2}{*}{6} & \multirow{2}{*}{3} \\
          &     &         &   $0.8, 0.915, 0.96$    &                       &                 &   & & \\
ecc$0.915$high&1.504 & $80.0$ & 0.915                      &  $-\omega x_c$         & 100           &0.125&258& 6 & 3 \\

\hline\hline
\end{tabular}
\end{table*}

\begin{figure}[t]
  \centering
    \includegraphics[width=0.49\textwidth]{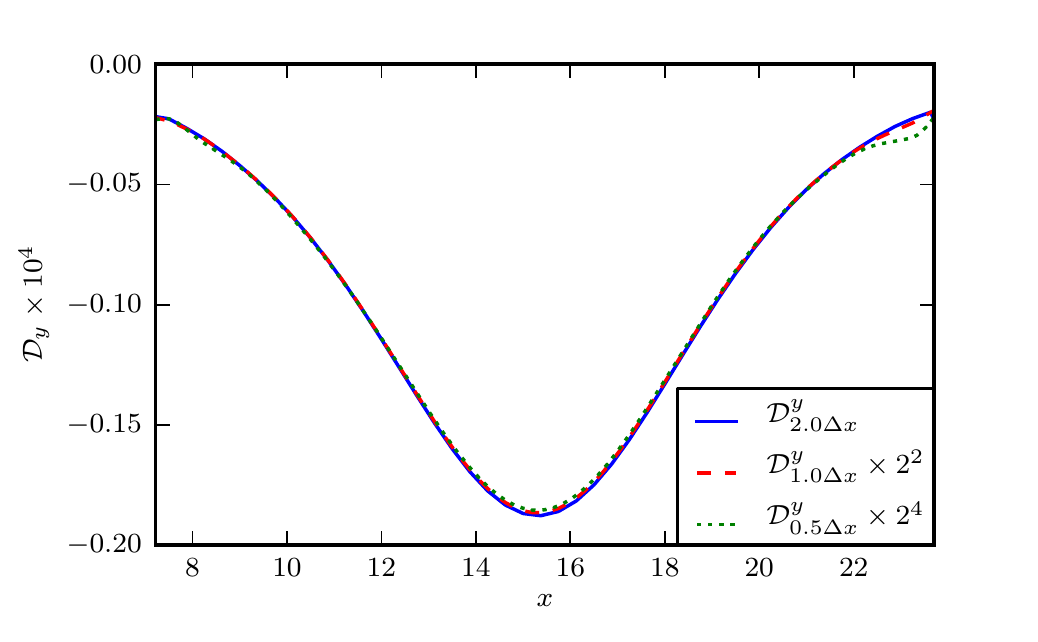}
    \caption{ \label{fig:momy_convergence} The $y$-component of the momentum constraint ($\cD^y$) for the  \emph{ecc0} setup. We plot $\cD^y$ along the $x$-axis (which passes through the centers of both stars) for three grid spacings of $\Delta x=0.1875$, $2\Delta x$, and $\Delta x/2$ in the finest box. The constraint violations were computed with second-order finite differencing, which represents the accuracy of the multigrid algorithm, so we scale the two finer resolutions as appropriate for second-order convergence. Note that the feature on the right side of the plot is due to inaccuracies at the surface of the star, leading to the spikes that we can also see in the eccentric case (see Sec.~\ref{s:eccentricconvergence})
    but do not show here to focus on the convergence in the strong-field interior of the star.
   }
    \end{figure}
First, we want to test convergence of the initial data solver using the \textit{ecc0} dataset. Therefore, we consider three different resolutions, with a finest grid spacing of $0.09375$, which is within the range of desired resolutions for production runs. We kept the outer boundaries fixed at a distance of $\sim 500$ with 5 levels of mesh refinement (where each level doubles the resolution), which is sufficient for our purposes.
One can use more refinement levels   for highly accurate data intended for  evolution and gravitational-wave extraction. Fig.~\ref{fig:momy_convergence} shows the expected second-order convergence in a one dimensional comparison of the momentum constraint. Here we plot the largest of the components, which is the $y$-component, since the stars are initially moving in the $\pm y$-direction. 
The convergence behavior for the other components or the Hamiltonian constraint is similar. At the surface of the stars we can see some deviation from  perfect convergence, including spikes in the constraint violations at the surface itself (shown in Sec.~\ref{s:eccentricconvergence} in the eccentric case), which are cut off here to show the central behavior in detail. These features are undesirable, but they are not easy  to remove in an implementation without surface-fitted coordinates.
The  convergence in the eccentric case is further detailed in Sec.~\ref{s:eccentricconvergence}, which includes a discussion of the spikes and a comparison of the convergence of the Hamiltonian and momentum constraints.

\begin{figure}[t]
  \centering
    \includegraphics[width=0.49\textwidth]{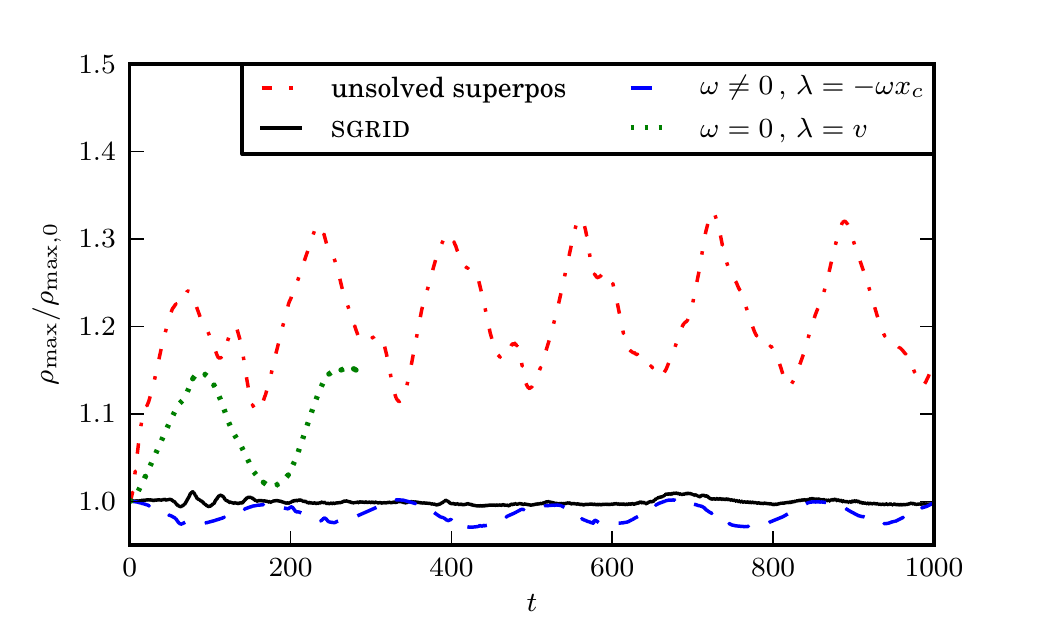}
    \caption{ \label{fig:densosc} Comparison of the oscillations of the star, measured using the maximum density at each timestep, $\rho_\text{max}$, normalized by the maximum density at $t=0$, $\rho_{\text{max, } 0}$. We show evolutions of \sgrid data (solid black) and the corresponding dataset constructed with our method assuming stationarity in a rotating frame (\textit{ecc0}, blue dashed). One can clearly see the improvement over the strongly oscillating curves of superimposed boosted spherical stars (red dot dashed) and data computed using stationarity in a linearly comoving frame (\emph{ecc0v}, green dotted). Note that the latter data set was only evolved for a short time, since we were only interested in the spurious oscillations.}
\end{figure}

As  mentioned earlier, using initial data inconsistent with the hydrodynamic
properties of the system can lead to spurious oscillations of the neutron
stars, which would contaminate the gravitational wave signal. As shown in
Fig.~\ref{fig:densosc}, the oscillations obtained in the evolution of the
\sgrid data (which solves for the velocity potential in addition to solving
the constraints) are negligible compared to those obtained when evolving
superimposed spherical TOV stars, which exceed $\sim 20\%$. The latter
configurations are generated by choosing a boost parameter that results in
an overall velocity that is similar to the known values for the
(approximately) quasicircular orbit from \sgrid and subsequently evaluating
the orbits and tuning the boost parameter to iteratively lower the orbital
eccentricity (via bisection).
Note that it is possible to reduce the spurious oscillations even for simple
superimposed TOV data by changing the stars' shapes, as found by Tsatsin and
Marronetti~\cite{TsaMar13}, who
adjusted the coordinates, matter density, and velocity in
an ad hoc but tunable way. This allowed them to reduce the oscillations by an order of
magnitude, even without explicitly solving the hydrodynamic and constraint
equations.

On the other hand, solving the hydrodynamic and constraint equations,
but assuming stationarity in a linearly comoving frame does not
significantly reduce the spurious oscillations.
As a noteworthy caveat, one should keep in mind that the configurations that
assume stationarity in a linearly comoving frame do not converge easily; one
must use significant softening (over-relaxation) and carefully adjust the
order in which the equations are iterated. Although the errors become
smaller and the solution seems to converge after a few iterations, the error
seems to saturate and the solution tends to diverge after a large number of
iterations if the error tolerance is small.
Since in this case, as discussed in Sec.~\ref{s:eccorbitshelsym}, true force
balance is lacking (a major source of instability, cf.~footnote 2), we did
not pursue this approach further.

By employing assumptions consistent with the hydrodynamic properties of the
system, i.e., assuming stationarity in a rotating frame at apoapsis
(discussed in Sec.~\ref{s:eccorbitshelsym}), the density oscillations
exhibited in simulations were reduced by an order of magnitude, i.e., to
$\sim 2$--$3 \%$, as shown in Fig.~\ref{fig:densosc}. It will be shown in
Sec.~\ref{s:eccentricconvergence} that the remaining density oscillations
can be further reduced (at approximately second order in our grid spacing)
by increasing the resolution. Unlike the method in~\cite{TsaMar13}, our
approach does not require any fine tuning, satisfies the constraint
equations, and leads to smaller density oscillations than the ad hoc method
in~\cite{TsaMar13}.

\section{Eccentric Orbits}
\label{s:ecc}

\subsection{Convergence}
\label{s:eccentricconvergence}

In this section, we perform evolutions of initial data sets constructed using the pair of inscribed helical symmetry vectors~\eqref{eq:hellipticalKV} for nonzero eccentricity. 
We first consider initial data for the \emph{ecc0.915(high)} cases (whose parameters are
given in Table~\ref{t:idpar}). The
grid setup is a realistic one that could be used for production-quality
evolutions, with $6$ levels of mesh refinement (beyond the coarsest grid level) and an outer boundary at
$\sim 10^3$. We consider two resolutions: The lower resolution has $130^3$ points in each
refinement level and a finest resolution of $0.25$ (with $64$ points across the star). The higher resolution has $258^3$
points in each refinement level and a finest resolution of $0.125$ (with $128$ points across the star). It took $0.6$ and $0.9$~hours to generate initial
data for these two configurations on the JUROPA cluster, running on $96$ and $256$ processors, respectively. This nicely illustrates the benefit of the simplicity of the implementation, as the ratio of time for evolution to initial data computation is satisfyingly large: Even for this very short run, the total wall clock time to merger (at $\sim 800$ simulation time) was $\sim 25$ and $\sim 50$ times longer than the time it took to solve for the initial data for the low- and high-resolution runs, respectively.
For longer evolutions this ratio is even larger [e.g., the total wall clock time to merger for the (low-resolution) \emph{ecc0.45} case is $\sim 6$ times longer than the \emph{ecc0.915} case, while the initial data solve took about the same time for both eccentricites].

\begin{figure}[t]
  \centering
    \includegraphics[width=0.49\textwidth]{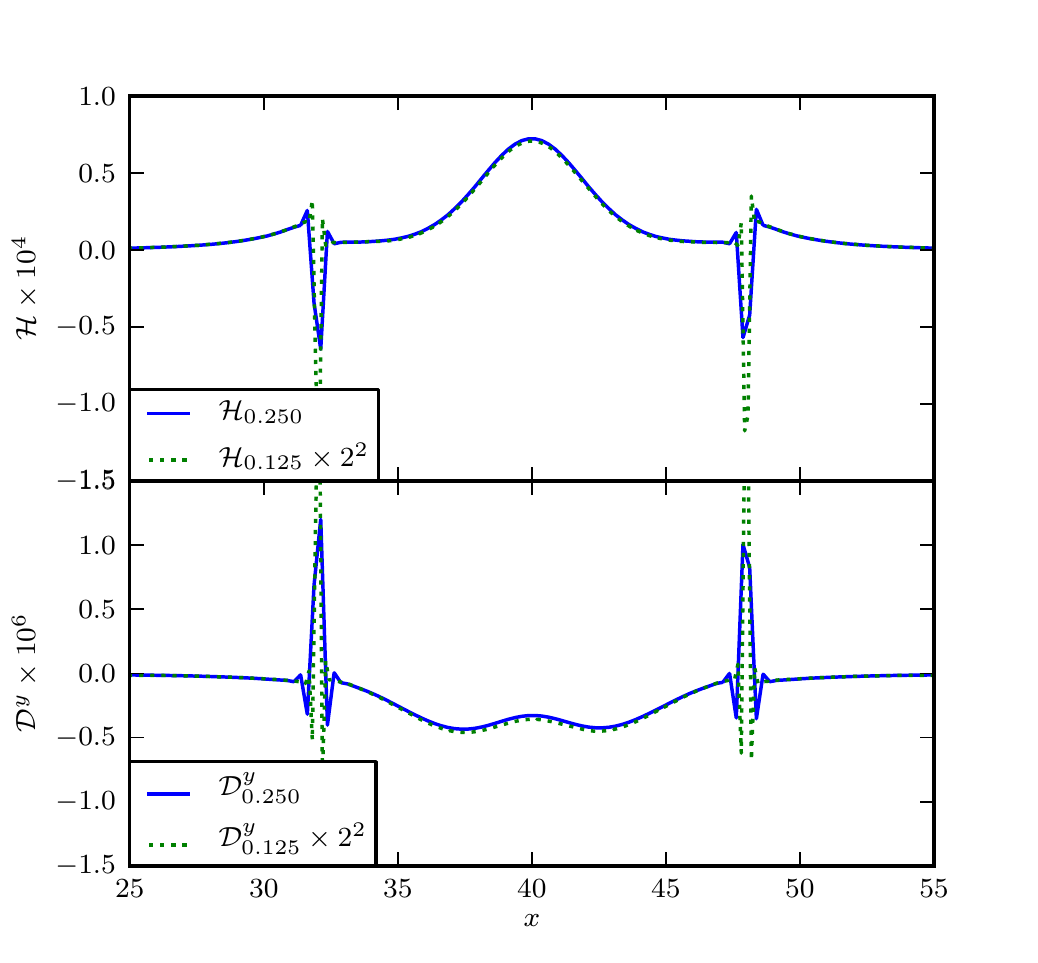}
    \caption{ \label{fig:e915_ham_conv_x6a} 
The initial constraint violations for the \textit{ecc0.915} data in the finest box
(surrounding one star) along the $x$-axis, which passes through both stars' centers.
We show the Hamiltonian constraint ($\cH$) and the $y$-component of the
momentum constraint ($\cD^y$) for two resolutions, with grid spacings of
$0.25$ and $0.125$ in the finest box and demonstrate 
second-order convergence of the constraints computed with sixth-order 
finite differencing stencils, the same order
used in the evolution of this dataset we show later. (Recall that the
initial data code is only second order accurate.)}
\end{figure}

\begin{figure}[t]
  \centering
    \includegraphics[width=0.49\textwidth]{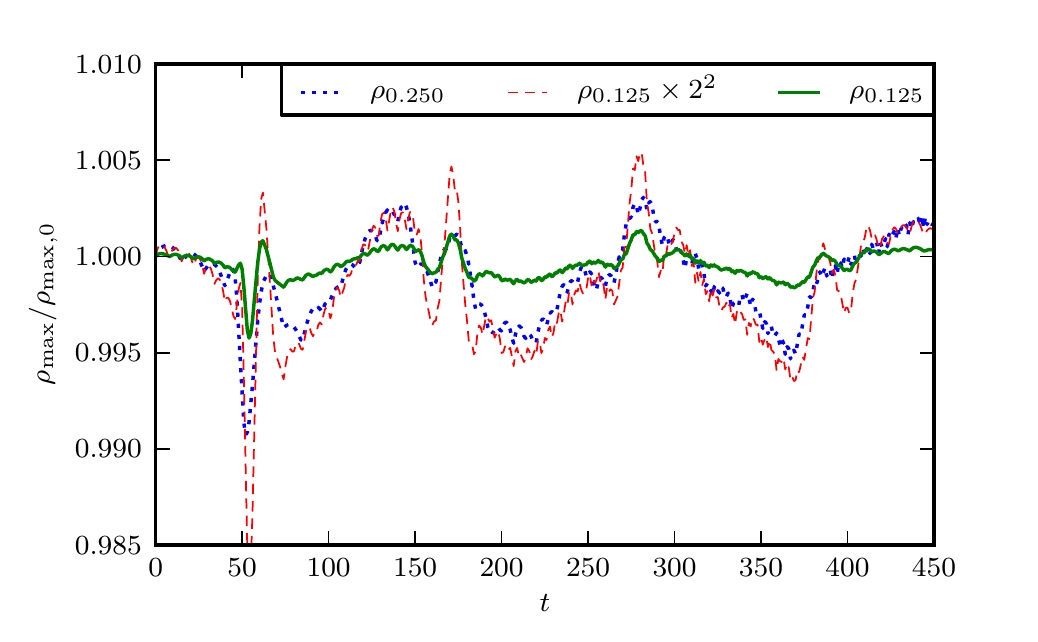}
    \caption{ \label{fig:convosc} The initial oscillations of a star in a binary with $e = 0.915$, illustrated by considering the maximum density at each timestep, normalized by the initial maximum density. We show this for two resolutions with finest grid spacings of $0.25$ and $0.125$, respectively (i.e., \emph{ecc0.915} and \emph{ecc0.915high}) to demonstrate how the density oscillations decrease with increasing resolution at approximately second order. The notation $\rho_{0.125}\times 2^2$ denotes that the oscillations (not the total maximum density) are multiplied by $2^2$, as discussed in the text.} 
\end{figure}


In Fig.~\ref{fig:e915_ham_conv_x6a}, we show convergence of the Hamiltonian and momentum constraints in the finest
box (surrounding one star) along the axis passing through both stars' centers. (We chose this dataset with a relatively large value of $e=0.915$ as a representative for all other eccentricities. The Hamiltonian and momentum constraint for other values of $e$ are almost the same as for the case at hand.) The convergence is clearly of second order, as expected, apart from some spikes at the surface of the star, which are also to be expected, since the density of the $n = 1$ polytropic stars we consider has a cusp at the surface (i.e., it is not differentiable there). Moreover, we have chosen to compute the constraints here using sixth-order finite differencing, the order we use in the evolutions (since the use of higher-order finite differencing improves their accuracy, e.g., reducing the constraint violations during the evolution), and one would expect the use of higher-order finite differencing to amplify any such features.

This increase of the finite difference order changes the shape of the constraint violations (cf.\ Fig.~\ref{fig:momy_convergence} and the bottom panel of Fig.~\ref{fig:e915_ham_conv_x6a}---there is not much difference due to the value of the eccentricity), in addition to changing their magnitude (increasing the
maximum magnitude at the center of the star for the Hamiltonian constraint, but decreasing it for the momentum constraint). This difference in shape is likely to 
be expected, since the remainders from second-order finite differencing are relatively large here. The use of higher-order finite differencing also creates spikes in the Hamiltonian constraint at the star's surface---there are only some slight wiggles present at the surface when the Hamiltonian constraint is computed using second-order finite differencing. However, increasing the order of finite differencing \emph{decreases} the size of the spikes at the surface of the star in the momentum constraint.

We also show the improvement of the spurious density oscillations with increased resolution in Fig.~\ref{fig:convosc}. We evolved the \emph{ecc0.915(high)} initial data described above (now using sixth-order spatial finite differencing, as discussed further in Sec.~\ref{ssec:trackswaves}) and monitored the maximum density. We see a clear improvement in the oscillations when doubling the resolution and the convergence can be estimated by multiplying the oscillations by the appropriate scaling factor, i.e., considering $4\Delta\rho_{\rm{max}}+\rho_{\rm{max,0}}$, where $\Delta\rho_{\rm{max}} = \rho_{\rm{max}}- \rho_{\rm{max,0}}$ for second-order convergence with a factor of $2$ difference in the grid spacing.
Apart from some smaller superimposed features, which can be seen, e.g., around $t=50$ or $t=180$, this scaling shows that the oscillations decrease with increasing resolution with almost second-order convergence. Of course, we do not expect the oscillations to completely converge away in the continuum limit, since we have still assumed spatial conformal flatness and have neglected the radial component of the velocity from radiation reaction.
However, we might expect that in the continuum limit these oscillations would be at the same small level seen for the \sgrid quasicircular data in Fig.~\ref{fig:densosc}, in which case it makes sense to compute the convergence order assuming that the oscillations are zero in the continuum limit. Indeed, this expectation is borne out by the results shown in the figure and preliminary results from our implementation of the method in {\sc{sgrid}}.

\subsection{Eccentric Sequences}
\label{s:eccentricseq}

As a check of our results, we compute constant-rest-mass sequences for equal mass stars of a fixed baryonic mass $m_b=1.625$ for varying eccentricity $e$ (the \emph{seq0}--\emph{seq0.9}
data sets in Table~\ref{t:idpar}); an isolated star with that baryonic mass has a gravitational mass of $M_i=1.5149$.  Given these quantities, we can compute the binding energy $E_b=M_{\mathrm{ADM}}-M$, where $M = 2 M_i$ and $M_{\mathrm{ADM}}$ denotes the Arnowitt-Deser-Misner (ADM) mass, an asymptotic quantity that gives a measure of the total mass of the spacetime. The ADM mass is defined via an integral at spatial infinity (see, e.g.,~\cite{OMuYor74}), and thus is generally obtained by extrapolation in
numerical codes that do not use compactified coordinates to include spatial infinity on the grid (see, e.g.,~\cite{BruGonHan06}). In our current situation, we found that the resolution of the outer grids was insufficient to allow us to obtain accurate results from extrapolation. We thus chose to obtain the ADM mass from a single sufficiently large extraction radius (though not too large, to avoid errors due to low resolution).
\begin{figure}[h!t]
  \centering
    \includegraphics[width=0.49\textwidth]{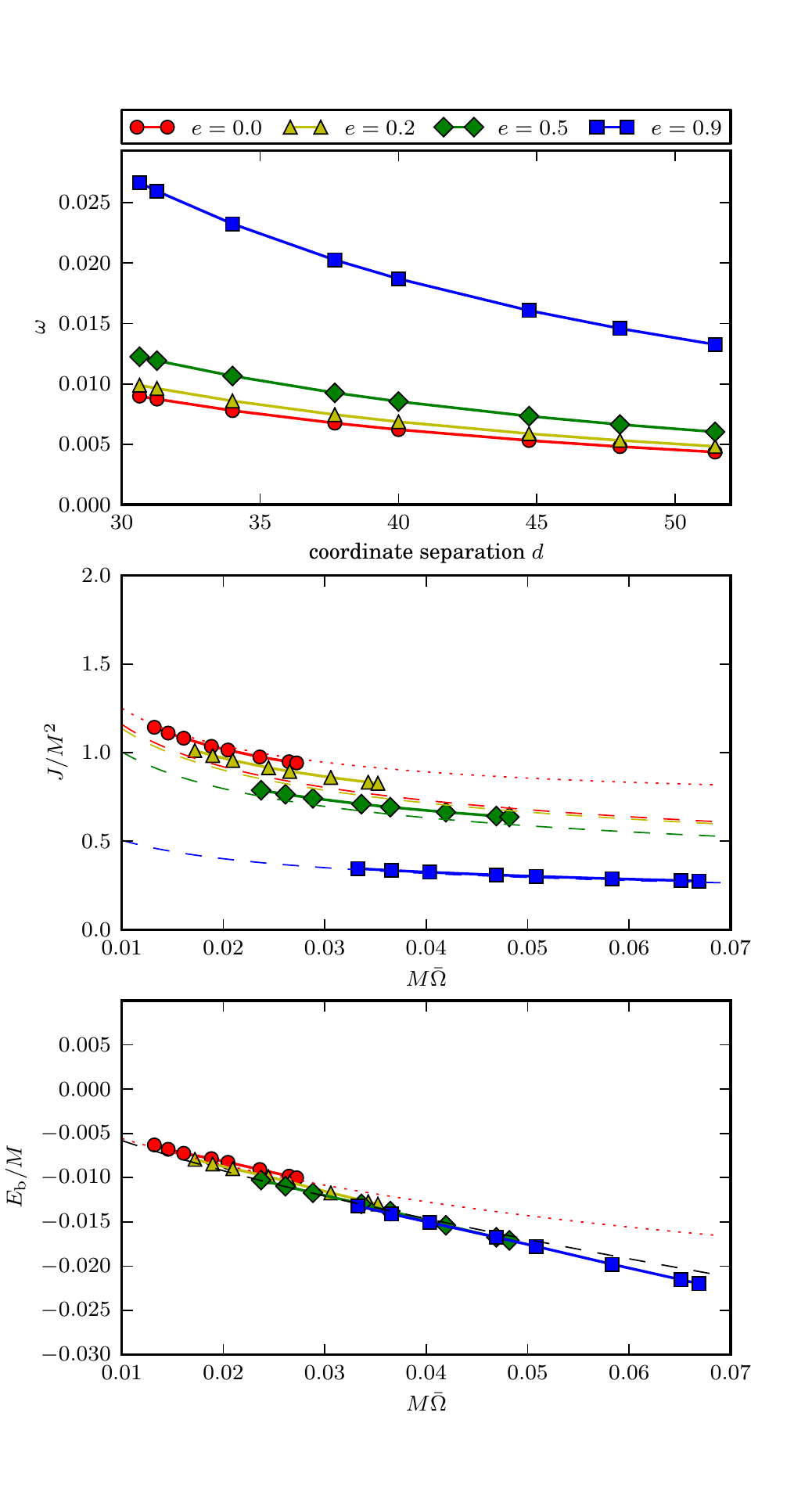}
    \caption{ \label{fig:seqEcc} Sequences for equal-mass binary neutron stars with varying eccentricity $e$ (data sets \emph{seq0}--\emph{seq0.9} in Table~\ref{t:idpar}). These sequences are computed with fixed baryonic masses, yielding isolated stars with gravitational masses of $M_i = 1.5149$; we define $M = 2M_i$. From top to bottom the quantities shown are the rotation $\omega$ as a function of the coordinate separation $d$ of the two stars' centers, and the ADM angular momentum $J_{\mathrm{ADM}}$ and binding energy $E_b = M_{\rm{ADM}}-M$ as functions of the normalized mean motion $M\bar\Omega$. The angular momentum has been normalized by $M^2$, while the binding energy is normalized by $M$. In addition, in the lower two plots, the expected Newtonian behavior is plotted in dashed lines (black line instead of color scheme for the bottom plot, since the Newtonian prediction is independent of $e$) and the post-Newtonian (3PN) results in dotted lines for $e=0$.
    }
\end{figure}

Here we can use a simplified formula for the ADM mass applicable to our spatially conformally flat case, given in Eq.~(16) of~\cite{OMuYor74}, which gives significantly better results with no extrapolation than the standard expression (given in, e.g., Eq.~(7) of~\cite{OMuYor74}). Specifically, in empirical tests with {\sc{sgrid}}  data, we found that changing the extraction radius from $r=150$ to $r=500$ leads to a $\sim4\%$ deviation in $M_{\rm{ADM}}$ when using the standard expression, while this deviation is less than $0.01\%$ when using the expression that takes advantage of conformal flatness. This is to be
expected, since (as discussed around Eq.~(16) in~\cite{OMuYor74}) the simplified expression can be evaluated at any radius in a region where the conformal factor satisfies the Laplace equation, and the conformal factor in our data satisfies the Laplace equation to a good approximation in the region in question [see Eq.~\eqref{eq:cts_ham2}], since the matter source is zero there, and the $A_{ij}A^{ij}$ term will be small (it falls off asymptotically as  the shift squared, and the shift goes to zero at infinity).
The standard measure for the ADM angular momentum is sufficiently accurate for finite radii, so that we do not use extrapolation here, as well (the deviation caused by the change of extraction radius considered above is around $0.05\%$).

For quasicircular data, equilibrium sequences are commonly supplied by plotting the dependence of the binding energy $E_b$ and ADM angular momentum $J_{\rm{ADM}}$  on the orbital frequency $\Omega$. In the eccentric case, we will use the mean motion $\bar\Omega$, rather than the frequency $\omega$ which appears  within the two inscribed helical symmetry vectors~\eqref{eq:hellipticalKV}. The reason  is that it is $\bar\Omega$, not $\omega$, that satisfies Kepler's third law \eqref{eq:Kepler3rd} in the Newtonian limit.
Specifically, increasing either $\omega$ or $\bar\Omega$ corresponds to decreasing the binary's initial coordinate separation, but only $\bar\Omega$ decreases with increasing eccentricity $e$, as expected from the Newtonian limit, while $\omega \propto 1/\sqrt{1-e}$ in the Newtonian limit. The top panel of Figure \ref{fig:seqEcc} gives a plot of $\omega$, as it is the quantity used in the code, which illustrates this increase
with increasing $e$.

Figure~\ref{fig:seqEcc} shows these sequences and compares them  to those expected from Newtonian theory. Considering the quasicircular limit $e=0$, we can see a qualitative agreement of the results. We also find the expected improvement in this agreement when we compare with the third order post-Newtonian (PN) curve (errors of $\sim 1\%$), based on calculations done by Mora and Will~\cite{MorWil03}. 
Considering the approximations we made for the velocity potential and the effect of finite size boxes on obtaining the ADM mass accurately, we do not expect better agreement.\footnote{While there are now $4$PN results available for the energy and angular momentum in the quasicircular case~\cite{Bla14}, which are given explicitly in Appendix~A of~\cite{KyuShiTan14}, we had initially intended to carry out the comparison with PN results in the eccentric case, as well, for which the $3$PN computations in Mora and Will~\cite{MorWil03} seemed the obvious choice.
Moreover, given the effects of our constant three-velocity approximation and finite boxes, comparisons with higher-order PN would not necessarily be too illuminating.} (See, e.g., Fig.~3 in~\cite{UryLimFri09} for comparison of a quasicircular sequence with $3$PN predictions using a code with surface-fitted coordinates that solves for the velocity potential.) 
One sees that the angular momentum decreases as the eccentricity increases in the middle panel of Fig.~\ref{fig:seqEcc}. This behavior is predicted by the Newtonian limit, as is the independence of the binding energy $E_b$ on the eccentricity seen (approximately) in the bottom panel of that figure.
Specifically, the Newtonian expressions in terms of $\bar\Omega$ are
\begin{subequations}
\begin{align}
\frac{E_b}{M} &= -\frac{\eta}{2}(M\bar\Omega)^{2/3},\\
\frac{J}{M^2} &= \eta(1-e^2)^{1/2}(M\bar\Omega)^{-1/3},
\end{align}
\end{subequations}
where $\eta := m_1m_2/M^2$ denotes the symmetric mass ratio, with $M$ the total mass, so $\eta = 1/4$ in the equal-mass case we are considering.
These expressions come from substituting $\Omega_a = [(1-e)/(1+e)^3]^{1/2}\bar\Omega$ into Eqs.~(2.2) in Mora and Will~\cite{MorWil03}. The expression for the angular velocity at aphelion, $\Omega_a$, in terms of $\bar\Omega$ comes from the standard Newtonian expressions above Eqs.~(2.2) in Mora and Will, noting that $\bar\Omega^2 = M/a^3$, where $a$ is the binary's semimajor axis, by Kepler's third law.

It would be obvious to compare these results to the PN calculations from Mora and Will, too. However, the PN results obtained in that manner are not well-behaved for head-on collisions. Since we approach such configurations for increasing eccentricities, while Mora and Will approach an unbound parabolic orbit (with a nonzero angular momentum), the comparison is not appropriate. 
While one could also consider the alternative quasi-Keplerian parametrization of eccentric orbits reviewed in Sec.~10 of~\cite{Bla14}, which remains well-behaved even for vanishing angular momentum through $1$PN, it is not clear how to relate the quantities used to describe the orbit in this parametrization to our $\bar\Omega$ variable using only data at apoapsis. Moreover, as noted by Sperhake~\emph{et al.}~\cite{SpeBerCar07}, the zero angular momentum limit is special even in the Newtonian limit, since all three types of orbits (elliptic, parabolic, and hyperbolic) degenerate to head-on collisions when the angular momentum vanishes, so it is perhaps not surprising that post-Newtonian results behave strangely there. (Sperhake~\emph{et al.}\ also give additional caveats about comparing eccentric post-Newtonian results with numerical relativity simulations.)

\subsection{Trajectories and waveforms}
\label{ssec:trackswaves}
\begin{figure}[t]
  \centering
    \includegraphics[width=0.49\textwidth]{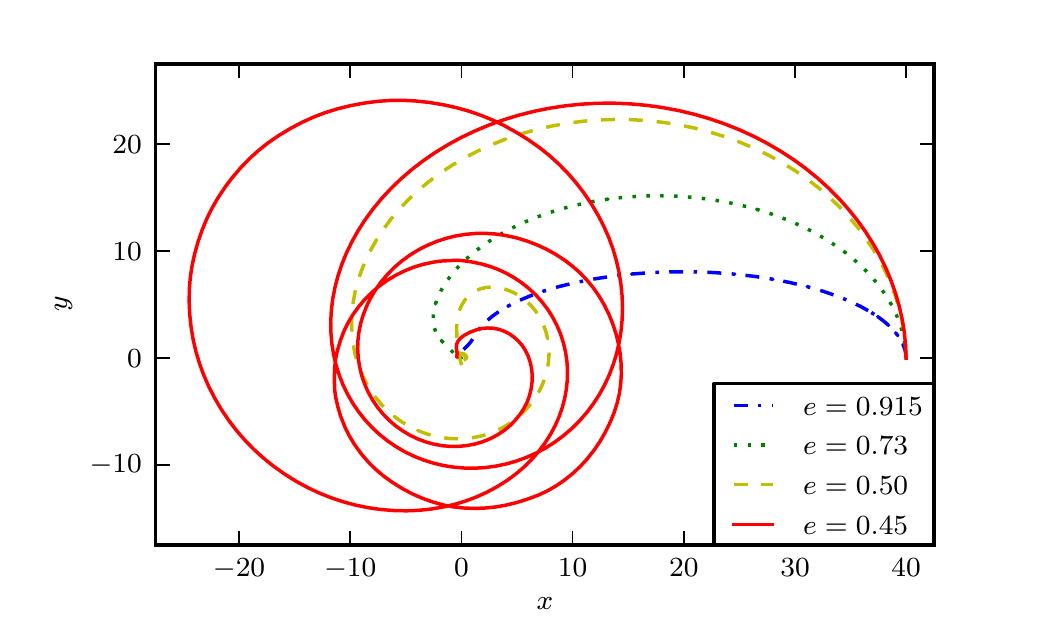}
    \caption{ \label{fig:ecctracks} The trajectories of one star for different values of the eccentricity parameter $e$. The evolutions are based on the initial data sets \emph{ecc0.45}, \emph{ecc0.5}, \emph{ecc0.73}, and \emph{ecc0.915} (see Table~\ref{t:idpar}) which are identical, except for the value of the eccentricity parameter $e$.  While larger values of $e$ lead to a rapid merger, for smaller values of the eccentricity (such as $e = 0.45$), one can obtain one or more encounters before merger, as found by Gold~\emph{et al.}~\cite{GolBerThi11}.
    } 
\end{figure}

To verify that the initial data obtained with our method actually show the desired features for nonzero eccentricities, it is useful to consider the star's trajectories. We define the trajectory of a star to be the coordinate position of the local minimum of the lapse at each time step. 
In Fig.~\ref{fig:ecctracks} we show that the eccentricity parameter $e$ has the expected dramatic influence on the orbits of the stars (as measured by their trajectories). We choose a series of initial data sets with fixed central enthalpy $h=0.255$, for which the gravitational mass of an isolated star is $1.399$, and fix the initial coordinate separation of the stars at $d=80$, but vary the eccentricity parameter. 
For the evolutions we again use the \bam code with the same settings given in Sec.~\ref{s:quasicirc}, except that we now use sixth-order spatial finite differencing to increase accuracy, as was done for binary black hole evolutions in Husa~\emph{et al.}~\cite{HusGonHan07}. Here we use eighth-order dissipation (with the same factor of $0.5$ used in the lower finite differencing order simulations), as is appropriate for sixth-order spatial finite differencing instead of Husa~\emph{et al.}'s choice of fourth-order dissipation (made for reasons of speed).

\begin{figure}[ht]
  \centering
    \includegraphics[width=0.49\textwidth]{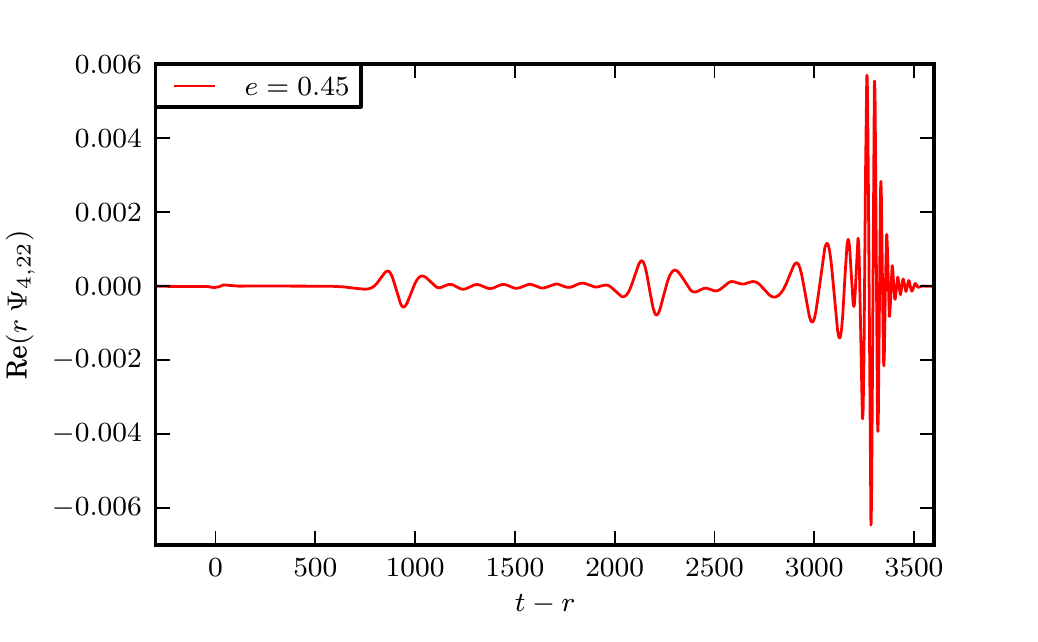}
    \caption{ \label{fig:eccwaves} The gravitational waveform (in the form of the $l = m = 2$ mode of the Newman-Penrose scalar $\Psi_4$) for the \emph{ecc0.45} case (see Table~\ref{t:idpar}). We extracted the waveform at a distance $r = 500$ from the binary's center-of-mass, and shift the time axis by $r = 500$ to account (approximately) for the waves' travel time. The two small bursts (at $t-r\simeq 900$ and $2100$)
    correspond to close encounters before merger; the tidally induced $f$-mode oscillations of the stars are visible inbetween the bursts.} 
\end{figure}

Since increasing $e$ yields a smaller tangential velocity of the star at apoapsis, the orbits become less circular, as the stars fall faster towards each other. In general, the number of orbits the stars perform before merger will decrease for eccentric orbits, since the configuration approaches a head-on collision as one increases $e$. However, it is possible to find interesting  configurations where the stars undergo one or more encounters before merger, as found in~\cite{GolBerThi11}. 
We illustrate this with an evolution with two encounters before merger (the \emph{ecc0.45} case) in Fig.~\ref{fig:ecctracks}, showing the gravitational waves for this case in Fig.~\ref{fig:eccwaves}. For the present 
illustration, we have chosen to present the waves extracted at
a finite radius, though we have checked that extrapolation to infinity (using radii from $300$ to $900$) only produces significant differences in the high-frequency part of the merger signal, where the grid spacing at the outermost extraction radii is likely too large to accurately transport waves with these high frequencies.

While it would seemingly be desirable to compare one or more of the trajectories shown in Fig.~\ref{fig:ecctracks} with a trajectory for the same case computed with superposed data (cf.\ the trajectories shown in Fig.~1 of Gold~\emph{et al.}~\cite{GolBerThi11}), we do not do so, since the comparison might actually be more confusing than illuminating. In particular, these are coordinate trajectories, and thus are a gauge-dependent quantity, and while the gauge conditions for both evolutions are the same, the initial gauge is not, since we initialize the lapse and shift in both cases using the values given by the initial data.
Indeed, the initial portions of the trajectories for the two cases look rather different, so while the qualitative zoom-whirl features in the evolution are the same as found for the superposed data, the initial portion of the trajectory for a run which has the same qualitative behavior of the trajectories and waveform looks much more eccentric with the superposed data than with the new data. (The Gold~\emph{et al.}~\cite{GolBerThi11} results also show more eccentric tracks for situations for which the trajectory and waveform have the same qualitative behavior, but here the primary difference is likely that the initial coordinate separation of the stars is almost twice as large as the one we consider here, while the masses are the same.)

The gravitational waveform shown in Fig.~\ref{fig:eccwaves} reveals the same key features found in Gold~\emph{et al.}~\cite{GolBerThi11}, specifically the high frequency signals between the bursts from pre-merger encounters: We have also checked that the frequency of these oscillations agrees with the $f$-mode frequency of an isolated nonrotating  star with the same baryonic mass and equation of state, as found by Gold~\emph{et al.} [We estimated this
frequency using the fits given in~\cite{LauLeuLin10} and the values of $1.399$ and $9.586$ for the isolated star's gravitational mass and areal radius; note that Eq.~(6) in~\cite{LauLeuLin10} contains a typographical error, which is corrected in Eq.~(14) of~\cite{ShaLinLeu14}.]

\begin{figure}[ht]
  \centering
    \includegraphics[width=0.49\textwidth]{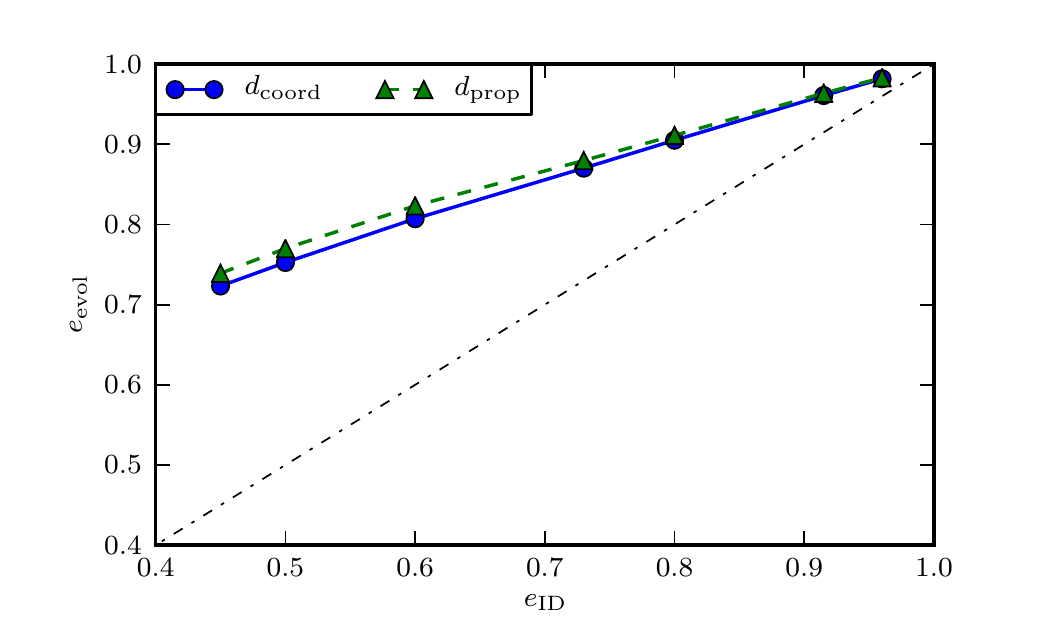}
    \caption{ \label{fig:eoutein} The relation of the eccentricity parameter $e_{\rm{ID}}$ input into the code (i.e., $e$ in previous plots) and the ``output'' eccentricity of the evolutions $e_{\rm{evol}}$ measured using two different fits to the initial portion of the trajectory, computed for the \emph{ecc0.45}--\emph{ecc0.96} data sets in Table~\ref{t:idpar}. The fit to an ellipse using the coordinate distance $d_\text{coord}$ gives quite similar results to the fit using the proper distance $d_\text{prop}$ and the trajectory's position angle. The black dot-dashed line shows the ``ideal'' relation $e_{\rm{ID}} = e_{\rm{evol}}$. The $d_\text{coord}$ and $d_\text{prop}$ curves agree well enough that one can use them to obtain an estimate of the eccentricity of the system being evolved. 
    }
\end{figure}
Since our method is only exact in the limit of Newtonian point particles, the eccentricity $e_{\rm{evol}}$ of the orbit obtained when evolving the initial data is not expected to be the eccentricity parameter $e_{\rm{ID}}$ used in the construction of the data (i.e., the parameter we have been referring to as $e$ so far). Therefore it would be useful to have a way to determine the eccentricity obtained in the evolution and relate this to the input eccentricity, allowing one to obtain a specific eccentricity, if desired. First, we will consider the different  methods available to determine $e_{\rm{evol}}$.
While there is no known definition of eccentricity in the comparable mass case in full general relativity, we can give a quantitative measure of the eccentricity of a given evolution by fitting an ellipse to a short section of the trajectory near the beginning of the orbit, as soon after the initial relaxation has
completed as possible. We expect that the trajectory will be most approximately elliptical there, since radiation reaction and other strong-gravity effects will not have had much time to affect the orbit. Besides fitting the track to an ellipse directly, it is also possible to measure the position angle of the trajectory and the proper distance between the stars and then fit an ellipse to the coordinates of the ``proper distance trajectory'' one obtains in this manner. This alternative method gives some indication of the extent to which the determination is contaminated by gauge effects (though the proper distance is not gauge invariant, since it is not computed along a geodesic). 
There are various other methods for eccentricity estimation in the literature (see the references cited in \cite{MroPfe12}, and also~\cite{KyuShiTan14} for more recent work), but most of them are only applicable to small eccentricities and also often require several orbits, so we do not consider them here.

Note that while the Newtonian definition of eccentricity we use, $e = (1 - b^2/a^2)^{1/2}$, seemingly requires knowing both the semimajor and semiminor axes of the orbit, $a$ and $b$, this is not actually the case. The method of performing a fit to a segment of the orbit that we use here does not require us to define either of quantities for the orbit as a whole---for which they would likely be undefined, or at least difficult to define, due to orbital precession---but only to obtain them for the ellipse that is the best fit to the segment of the orbit we consider, where they are given directly by the fit. Obviously, the resulting eccentricity estimate should merely be taken as a reasonable way of measuring the eccentricity which gives a qualitative feel for how eccentric the orbit is, rather than anything fundamental. In particular, since this estimate uses the trajectories, which are gauge-dependent, one might obtain very different results for this estimate if one evolved the same data using a code which used substantially different gauge conditions, e.g., the Princeton group's code, which uses generalized harmonic coordinates, and has been used to study highly eccentric systems~\cite{EasPre12,SteEasPre11,EasPreSte12}.

We give the relation between the
eccentricity in the evolution we measure using these methods ($e_{\rm{evol}}$) and the eccentricity parameter input to the code ($e_{\rm{ID}}$) in Fig.~{\ref{fig:eoutein}}. From
experimentation with the interval over which we fit the ellipse, we can estimate the error introduced by using different intervals for the fit (varying both the location and size of the interval). Specifically, we vary the interval length between 200 and 3000 points, which corresponds to coordinate displacements from $\sim 0.8$ to $\sim 9$ or a evolution timespan of $\sim 30$ to $\sim 200$, and see that the deviations are $\sim 1\%$, or even smaller if one just considers medium sized intervals that do not cover the very first points. If we compare the eccentricities computed using the coordinate and proper distances, we find results that agree within 3\% for eccentricities larger than $0.4$.
This agreement is sufficient for the purpose of constructing eccentric orbits, since it merely serves as a rough estimate of the expected eccentricity obtained in the evolution of the data. Note also that we have computed Fig.~{\ref{fig:eoutein}} for the same choice of stellar masses as in Fig.~\ref{fig:ecctracks} (i.e., using the \emph{ecc0.45}--\emph{ecc0.96} data sets in Table~\ref{t:idpar}), but the results one obtains for different stellar masses are very similar. In particular, a reduction of $\sim10\%$ in the mass only resulted in a change of $\sim0.5\%$ in $e_{\rm{evol}}$, independent of the method used to compute it.

Additionally, note that the method we have used to determine the 
eccentricity is only applicable for $e\gtrsim0.5$. For small $e$, one obtains inaccurate results due to radiation reaction. In particular one sees both a larger deviation of the eccentricity measured with the coordinate and proper distances for small $e$, as well as a noticeable offset (of $\sim 0.3$) for quasicircular data (though this offset is smaller than the value of $\sim 0.5$ that would be predicted by linearly extrapolating the curve in Fig.~{\ref{fig:eoutein}} back to $e = 0$). Such a large offset is not seen when one uses any of the (previously mentioned) eccentricity determination methods that are specialized to small eccentricities.

Altogether, Fig.~\ref{fig:eoutein} shows that the eccentricities we obtain in evolutions behave as expected (i.e., increase monotonically as $e_\text{ID}$ increases). 

\section{Summary and outlook}
\label{s:concl}

There are certain scenarios in which binary neutron stars can merge without having shed all of their eccentricity, e.g., due to dynamical capture, and simulations in general relativity are the only way to model
such mergers accurately (necessary to study, e.g., their gravitational waves, ejecta, and merger remnants). In this paper we have given the first method capable of providing consistent initial data for such systems (i.e., initial data that solves both the constraint equations of general relativity and the Euler equation). Our method proceeds by generalizing the approximate helical Killing vector that  is used to solve the Euler equation via its first integral in the quasicircular case to a pair of inscribed helical symmetry vectors (one for each star), which allows us to provide initial data for binary neutron stars with arbitrary eccentricity. We find that the initial spurious oscillations found in evolutions of inconsistent data are reduced by an order of magnitude or more (with higher resolution) using our consistent data, which assumes stationarity in a rotating frame, while the spurious oscillations remain if one assumes stationarity in a linearly comoving frame. We also find that the oscillations induced by the tidal deformation at each close encounter are indeed physical and
not qualitatively altered compared to those observed in earlier  evolutions of  initial data that did not solve the Euler equation~\cite{GolBerThi11,EasPre12}.

Considering the foundations of the method, we give two motivations for the inscribed helical symmetry vectors we introduce. In one derivation, we add a boost to the standard approximate helical Killing vector used for quasicircular initial data to adjust the binary's velocity at apoapsis, allowing one to control the binary's eccentricity. In the other, more geometrical derivation, we show how the same vector arises from approximating an elliptical orbit at apoapsis using an inscribed circle. We also show that the fact that there are two different inscribed helical symmetry vectors (one for each star) does not spoil the derivation of the extended conformal thin-sandwich equations, by showing that one can obtain these equations without assuming Killing symmetry by imposing spatial conformal flatness and maximal slicing from the outset. Additionally, we give a further extension of the method that allows one to add radial velocity to the stars, so that (at least in principle) one can obtain consistent binary neutron star initial data with arbitrary initial tangential and radial velocities. 

For our first implementation of the method we introduced, we chose to use a
second-order Cartesian multigrid solver (i.e., without surface-fitted
coordinates), for simplicity and comparative speed. 
Without surface-fitted coordinates,
one cannot easily solve for the velocity potential, so we have taken the
$3$-velocity to be constant, which we show is a good approximation if the
stars are not too close. However, this is not a requirement of the method,
and one can easily solve the equation for the velocity potential, as well,
if one is using a code that employs surface-fitted coordinates (e.g.,
{\sc{sgrid}}~\cite{Tic09a}).

There are many potential extensions of the method, as discussed in the paper, and we are already in the process of implementing and testing a number of them. A first and straightforward step will be to construct binaries with unequal mass stars. In addition, we are currently extending the {\sc{sgrid}} code to use our inscribed helical symmetry vectors, in order to take advantage of {\sc{sgrid}}'s spectral accuracy and surface-fitted coordinates, which would allow us to solve for the velocity potential easily. 
The \sgrid implementation will also naturally allow us to use more realistic EOSs, modeled as piecewise polytropes, which \sgrid has recently been extended to handle. Furthermore, this implementation should make it possible to add arbitrary spin to the stars, which might be especially pertinent for eccentric systems, as discussed in Sec.~\ref{s:intro}. 

We have also implemented the generalized vector that includes radial velocity and now need to investigate the properties of the data we obtain from it. In particular, since the generalized vector allows one to modify both the radial and tangential velocity components of the stars, it should allow us to obtain low-eccentricity initial data, similar to the work done for black hole-neutron star binaries in~\cite{FouKidPfe08} and very recently for binary neutron stars in~\cite{KyuShiTan14}, or the various well-established methods for eccentricity reduction for binary black holes~\cite{PfeBroKid07,HusHanGon07,TicMar10,MroPfe12,PueHusHan12}.
Low-eccentricity binary neutron star initial data are particularly important from a gravitational wave data analysis point of view: The residual eccentricity in current simulations is large enough to bias determination of the tidal deformation~\cite{ReaBaiCre13}, which would provide a valuable constraint on the poorly-known equation of state of cold, dense nuclear matter. 

It may also be interesting to consider PN corrections to the Newtonian expressions for the orbital motion used in deriving the inscribed helical symmetry vectors in order to obtain sequences that yield better agreement with PN predictions. This would also facilitate  comparisons with analytic techniques, such as PN or EOB formulations with tidal corrections and eccentricity.
We note, however, that neglecting PN corrections to the orbital motion in our  construction mostly affects the relation between the value of the eccentricity parameter $e$ used in the code and the eccentricity obtained in the simulation, and not the accuracy of the initial data or simulation: One can always iterate over $e$ to obtain any desired eccentricity in the simulation.

Even without the  possible extensions of the method, one may already make certain  useful investigations with the current initial data. In particular, in the near future, we intend to  revisit and extend the studies of Gold~\emph{et al.}~\cite{GolBerThi11} and East and Pretorius~\cite{EasPre12} (e.g., concerning properties of the merger remnant and ejecta) in order to determine  qualitative and quantitative  changes  upon using improved initial data, 
before going on to study more general scenarios (considering eccentricity reduction, adding spin, etc.).
The ability to construct self-consistent initial data for eccentric binary neutron stars  opens the door to studying many interesting physical situations, in both the high- and low-eccentricity regimes, without significant limitations in  accuracy.



\acknowledgments
 We are grateful to  K\=oji Ury\=u for his guidance and for pointing out that initial data can be constructed without assuming equilibrium if the IWM approximation and maximal slicing are employed. We also thank Sarp Akcay, Leor Barack, Thomas Baumgarte, Sebastiano Bernuzzi,  John L.\ Friedman, Eric Gourgoulhon, Alexandre Le Tiec, Marcus Thierfelder, and Clifford Will for fruitful discussions and suggestions. Additionally, we thank the anonymous referee for useful comments.  We gratefully acknowledge  support from the DFG SFB/Transregio 7 
``Gravitational Wave Astronomy'', DFG
Research Training Group 1523/1 ``Quantum and Gravitational Fields'', STFC grant PP/E001025/1, NSF grant PHY-1305387,
and the supercomputing grant  from the John von Neumann Institute for Computing (NIC)  provided on  JUROPA at the J{\"u}lich Supercomputing Centre (JSC).

\appendix*
\section{Injection energy and velocity potential for nonrelativistic incompressible binaries}  \label{sec:EulerFinsler} 

The nonrelativistic Euler equation for barotropic flows can be written in the Crocco  form \cite{Gou06} 
\be \label{eq:a1}
 {\partial _t}{v_a} + {v^b}(\bar{D}_b{v_a} - {\bar{D} _a}{v_b}) =- {\bar{D} _a}H,
\ee
where $H=\frac{1}{2}v^2+h_{\rm N}+\Phi$ is the Hamiltonian of a fluid element with specific enthalpy $h_{\rm N}$, 
$v^a = d x^a/dt$ is its velocity, $\Phi$ is the gravitational potential, and $\bar{D}_a$ is the covariant derivative compatible with the Euclidian 3-metric $f_{ab}$ in $\mathbb{E}^3$. (As before, we have used the same
notation for this covariant derivative as for the covariant derivative compatible with the conformal $3$-metric, since we took the conformal $3$-metric to be flat.) If the flow is irrotational, $v_a = \bar{D}_a \Psi$, then Eq.~\eqref{eq:a1} has a first integral
\be \label{eq:a2}
\partial_t \Psi =-H.
\ee
For incompressible flows, the specific enthalpy is given by 
$h_{\rm N}=\int dp/\rho=p/\rho$, where $p$ is the pressure, and
the mass density $\rho$ is constant.
Then, the continuity equation
\be \label{eq:a3}
\partial_t \rho+\bar{D}_a(\rho v^a)=0
\ee
simplifies to a Laplace equation for the velocity potential,
\be \label{eq:a4}
\bar{D}_a \bar{D}^a \Psi=0.
\ee
In what follows, we construct analytic solutions to 
Eqs.~\eqref{eq:a2} and \eqref{eq:a4}
for  binaries on circular and eccentric orbits.

\subsection{Circular orbits}
For irrotational incompressible binaries on circular orbits, all fluid elements  move on circles with different centers, but with the same radius $R$, and with the same speed $v=\Omega R$. Then, Eqs.~\eqref{eq:a2} and \eqref{eq:a4} have the exact solution
 \be \label{eq:a5}
\Psi(t,{\vv{r}}) =  - {\cal E}t + {\vv{v}}(t) \cdot {\vv{r}} =  - {\cal E}t - \Omega R(x\sin \Omega t - y\cos \Omega t)
\ee
where $\cal E$ is the injection energy (which is constant in both space and time)~\cite{FriedmanStergioulas2013},
\be \label{eq:a6}
\vv v (t)= \vv{\dot R}=- \Omega R(\sin \Omega t \, \vv{\hat x} - \cos \Omega t  \, \vv{\hat y})= \vv{\nabla} \Psi
\ee
is the fluid velocity and 
\be \label{eq:a7}
\vv{R}(t)=R \cos \Omega t   \,\vv{\hat x} + R\sin \Omega t  \, \vv{\hat y}
\ee is the position of the star's center relative to the center of mass (assumed here to coincide with the origin). Here we use $\vv{\nabla}$ for the index-free version of $\bar{D}_a$. It is straightforward to check that 
\be \label{eq:a8}
({\partial _t} + \Omega {\partial _\varphi })\Psi =[{\partial _t} + \Omega(x\, {\partial _y -y {\,\partial _x }})] \Psi =- {\cal E}.
\ee
This equation can be interpreted as Eq.~\eqref{eq:a2} transformed to a rotating frame. Alternatively, it may interpreted as a first integral of the equation
\be \label{eq:a9}
({\partial _t} + \Omega {\Lie{\vv \varphi}}){v_a} = 0,
\ee
which follows from helical symmetry, i.e., stationarity in a rotating frame.
The conserved injection energy follows from Eqs.~\eqref{eq:a2} and \eqref{eq:a8} and reads
\be
\mathcal{E}=H-\Omega (xv_y-y v_x).
\ee
Taking the gradient of this equation and evaluating at the center of the star $(R,0,0)$ at $t=0$  gives a force balance equation
\be
\partial_x \mathcal{E}|_{x=R}= \partial_x \Phi|_{x=R}-\Omega^2 R =0,
\ee
which yields Kepler's third law for inverse square forces.\\

\subsection{Eccentric orbits}
We wish to generalize this derivation to eccentric binaries. In this case, the position of the stellar center relative to the center of mass (assumed again to coincide with the origin) is  given by
\be \label{eq:a12}
{\vv{R}}(t) = [a\cos \zeta (t) + ae]{\vv{\hat x}} + b\sin \zeta (t) {\vv{\hat y}},
\ee
where $\zeta(t)$ is the eccentric anomaly, related to the mean anomaly $\bar\Omega t$ via the Kepler equation
\be \label{eq:a13}
\bar\Omega t = \zeta (t) + e\sin \zeta (t).
\ee
Here, $a$, $b$, $e$ and $\bar\Omega$ are the semi-major axis, semi-minor axis, eccentricity, and mean motion of the orbit of one star, respectively.
For simplicity, we  have chosen to study an effectively one body problem by assuming an extreme mass ratio, so the other, massive star (and thus the center of mass) is at the origin, which is chosen to be  the left focus of the ellipse. It is straightforward to relax the extreme mass ratio assumption and recover the two body equations, but  we defer this until  the end of this section. 
We have also assumed that the values $\zeta=0$ and $\zeta=\pi$
correspond to apoapsis and periapsis, respectively. The Kepler equation \eqref{eq:a13} has a series solution
\be \label{eq:a14}
\zeta(t) = \frac{{\bar\Omega t}}{{e + 1}} + \frac{{{{(\bar\Omega t)}^3}e}}{{6{{(e + 1)}^4}}}   + \mathcal{O}({t^5}).
\ee
The fluid velocity is  homogeneous and given by
\be \label{eq:a15}
\vv v (t)= \vv{\dot R}=- \dot \zeta(t)[a \sin \zeta (t) \, \vv{\hat x} - b \cos \zeta (t)  \, \vv{\hat y}]=\vv{\nabla} \Psi
\ee
and the velocity potential is given by
\be
\begin{split}
\Psi(t,{\vv{r}}) &=  - {\cal E}t + {\vv{v}}(t) \cdot {\vv{r}}\\
&=  - {\cal E}t - \dot \zeta (t)[a \, x\sin \zeta (t) - b \,y\cos \zeta (t)].
\end{split}
\ee
\\
If we operate on the above expression with
${{\partial _t} +  k^i }{\partial _i } = {\partial _t} + \omega {\partial _\varphi } + \lambda {\partial _y}$, where $k^i$ is the spatial part of our inscribed helical symmetry vector \eqref{eq:hellipticalKV}, and demand that the resulting expression be constant throughout the star at $t = 0$, i.e., $\vv{\nabla}{\cal E} = 0$, we obtain 
 \be \label{eq:vomegae2}
v = (1-e^2)\, \omega \, a
\ee
with $\omega$ given by Eq.~\eqref{eq:omegaeOmega}. If, in addition, we demand that
$k^i=v y^i$ at the star center $x=a(1+e)$ at $t=0$, we obtain
Eq.~\eqref{eq:lambdave}.
It is straightforward to check that 
the force balance equation 
\be
\partial_x \mathcal{E}|_{x=a(1+e)}=0
\ee
applied to the star center  for inverse square forces yields Kepler's third law  for eccentric binaries. 
To recover the two-body equations, it suffices to rescale the ellipse 
by a factor depending on the mass of each companion, as indicated by Eq.~\eqref{eq:centersx1x2}. Then,
Eq.~\eqref{eq:vomegae2} is replaced by Eqs.~\eqref{eq:vomegae}, while $e$, $\omega$, and $\bar\Omega$ remain unchanged.

\bibliography{Refs/refs,Refs/library}

\end{document}